\documentclass[fleqn,usenatbib]{mnras}
\usepackage{newtxtext,newtxmath}
\usepackage[T1]{fontenc}
\usepackage{ae,aecompl}
\usepackage{hyperref}

\usepackage{graphicx}	% Including figure files
\usepackage{amsmath}	% Advanced maths commands
\newcommand{\lya}{Ly-$\alpha$}
\newcommand{\lyb}{Ly-$\beta$}
\newcommand{\ovi}{\ion{O}{VI}}
\newcommand{\civ}{\ion{C}{IV}}
\newcommand{\hi}{\ion{H}{I}}
\newcommand{\heii}{\ion{He}{II}}
\newcommand{\SiIII}{\ion{Si}{III}}
\newcommand{\SiIV}{\ion{Si}{IV}}
\newcommand{\lymana}{Lyman-$\alpha$}

\newcommand{\kms}{ km s$^{-1}$ }

\newcommand{\Secref}[1]{\hyperref[#1]{Section~\ref{#1}}}

\defcitealias{Pieri2014}{P14}
\defcitealias{Pieri2010Stacking}{P10b}
\newcommand{\Morrison}{Morrison et al. in preparation}

\title[Probing UV Background Inhomogeneity with Quasars]{Probing Large-scale UV Background Inhomogeneity Associated with Quasars using Metal Absorption}

\author[Morrison et al.]{
Sean Morrison,$^{1}$\thanks{E-mail: sean.morrison@lam.fr}
Matthew M. Pieri,$^{1}$\thanks{E-mail: matthew.pieri@lam.fr}
Debopam Som,$^{1,2,3}$
and Ignasi P\'{e}rez-R\`{a}fols$^{1,4}$
\\
$^{1}$Aix Marseille Universit\'{e}, CNRS, CNES, Laboratoire d'Astrophysique de Marseille, Marseille, France\\
$^{2}$Department of Astronomy, The Ohio State University, 140 West 18th Avenue, Columbus, OH 43210, USA\\
$^{3}$Space Telescope Science Institute, 3700 San Martin Drive, Baltimore, MD, 21218, USA\\
$^{4}$Sorbonne Universit\'{e}, Universit\'{e} Paris Diderot, CNRS/IN2P3, Laboratoire de Physique Nucl\'{e}aire et de Hautes Energies,\\ LPNHE, 4 Place Jussieu, F-75252 Paris, France
}

\date{Accepted XXX. Received YYY; in original form ZZZ}

\pubyear{2020}

\begin{document}
\label{firstpage}
\pagerange{\pageref{firstpage}--\pageref{lastpage}}
\maketitle

\begin{abstract}

We study large-scale UV background inhomogeneity in three-dimensions associated with the observed quasar populations at high redshift. We do this by measuring metal absorption through quasar absorption spectrum stacking as a function distance to closest quasar in SDSS-IV/eBOSS on 10s of comoving megaparsec scales. We study both intergalactic medium absorbers and mixed circumgalactic medium absorbers and probe absorption in \ovi, \civ, and \SiIV, and \SiIII. Overall stronger high ionization species absorption is seen closer to quasars at $2.4<z<3.1$. \ovi\ absorption shows a particularly strong change, with effects in \civ\ evident in some cases, and more marginal effects in \SiIII\ and \SiIV. We further study $2.05<z<2.4$ (with weak signs of increasing homogeneity with time) and explore the study of metal absorption as a function of integrated SDSS-r band flux quasar flux (yielding consistent but less significant results). While the metal absorption does show sensitivity to large-scale 3D quasar proximity, the current incomplete quasar samples limit detailed interpretation. This work does, however, demonstrate that UV background inhomogeneities exist on scales of several 10s of comoving megaparsecs associated with quasars and that they can be measured with precision by examining metal absorption in the intergalactic medium.

\end{abstract}


%%%%%%%%%%%%%%%%%%%%%%%%%%%%%%%%%%%%%%%%%%%%%%%%%%%%%%%%%%%%%%%%%%%%%%%%%%%%%%%%%%%%%%%%%%%%%%%%%%%%
%%%%%%%%%%%%%%%%%%%%%%%%%%%%%%%%%%%%%%%%%%%%%%%%%%%%%%%%%%%%%%%%%%%%%%%%%%%%%%%%%%%%%%%%%%%%%%%%%%%%
%%%%%%%%%%%%%%%%%%%%%%%%%%%%%%%%%%%%%%%%%%%%%%%%%%%%%%%%%%%%%%%%%%%%%%%%%%%%%%%%%%%%%%%%%%%%%%%%%%%%

%%%%%%%%%%%%%%%%% BODY OF PAPER %%%%%%%%%%%%%%%%%%%%%%%%%%%%%%%%%%%%%%%%%%%%%%%%%%%%%%%%%%%%%%%%%%%%
\section{Introduction}
Our knowledge of the properties of the intergalactic medium (IGM) and the circumgalactic medium (CGM) inform our understanding of the joint evolution of galaxies and their environments. Galaxies deposit energy in the environments, both mechanically and radiatively, modifying gas metallicity bulk motion, thermal motion and ionization. Since the accretion of this gas drives star formation and active-galactic nucleus (AGN) activity, the study of this energy deposition is not purely diagnostic; it enables us to understand subsequent feedback in these systems. The low densities typical of the high redshift ($z\gtrsim 2$) IGM make it optically thin to ionizing photons, while the moderate temperatures ($10^4$K $< T < 10^5$K) result in minimal collisional ionzation. It is, therefore, thought to be dominated by the raditive impact of emitted UV photons from the so-called UV background. 

The ionizing photons that permeate the IGM result in a extergalactic background that both reflects the global properties of the Universe and help to drive them. Different spectral ranges of the background originate from different sources. The UV background is produced by quasars (via AGN) and galaxies (via stars). The properties of stars and galaxies are the focus of various observations and models \citep{Dominguez2011, Lagos2019}. However, the sample size of observations, especially at high-$z$, limit our knowledge of their properties. Their contribution to the UV background remains poorly understood due to factors such as the uncertainties in the escape fraction of photons from the galaxies \citep{Shapley2006}. Additionally, the lifetime and duty cycle of quasars are also not well known. Passive quasars will leave a fossil zone, where light travel time and metal recombination times leave a detectable imprint \citep{Oppenheimer2013} on metal abundances and the UV background. Despite knowing that there are fossils zones, the proportion of time a quasar is active or passive remains an unknown. 

Since photons associated with the UV background are produced from discrete sources, the UV background is not homogeneous at all scales and for all frequencies. Around individual sources, there are regions of excess ionizing photons. These regions are known as proximity zones. The size and filling fraction of these regions depends on the photon escape fraction of the surrounding gas, the lifetime (or duty cycle) of the source, and the mean free path of photons. The contribution from quasars is complicated by the effect such as the degree of quasar beaming, light echos \citep{Visbal2008}, and AGN duty cycles \citep{Martini2001}. Light travel time can be probed by studying the difference between foreground and background quasar proximity regions.

The mean free path of photons is a function of density and ionization state of the material, as well as the energy of the photons. The mean free path of ionizing photons has evolved through time. During \hi\ reionization, the mean free path (determined from simulation) evolves from $\lesssim 10$~Mpc at $z\gtrsim 10$ to $\gtrsim 10$~Mpc at $z\lesssim 10$ \citep{Rahmati2018}. By $z\sim 2-3$, the mean free path of \hi\ ionizing photons has increased to $\sim100-300~ h^{-1}$ comoving megaparsec (cMpc) \citep{Croft2004, Rudie2013}, which is larger than the mean separation of the sources. Other factors also contribute to the scales of inhomogeneities. \citet{Wyithe2008} included the bias of reionization around quasars into a model for evolution of the ionization of the IGM and the emissivity of ionizing photons. They predict that at high redshift, before \hi\ reionization, quasars ionized local IGM regions of $\sim 5$~Mpc well before ($\delta z\sim 0.3$) the typical IGM. This, in combination with the rapid increase in the mean free path of ionizing photons expected when the regions overlap, results in an intensified UV background near the quasars. They also found that, post reionization, the size of the regions is related to the surrounding IGM and the background ionization rate.

At $z\sim 3$, a phase change occurs due to \ion{He}{II} reionization \citep{Shull2010,Syphers2011,Worseck2016} altering the UV background and heating the IGM \citep{Miralda-Escude1994}. While the scales would most likely differ between \hi\ reionization ($z\sim6$) and \heii\ reionization, the properties will be similar. The mean free path of \heii\ ionizing photons is significantly shorter than for \hi\ ionizing photons during the end, and immediately following, the epoch of \heii\ reionization. Estimates of the \heii\ mean free path at $2.5<z<3.2$ have been approximated to scales of $20-100$~cMpc, with the range due to the numerous factors involved in its determination (e.g. \citealt{Furlanetto2009,Davies2014,Faucher-Giguere2009,Davies2017}). Additionally, \citet{Davies2017} found a spatially varying mean free path, that contributes to UV background fluctuations on scales up to $\sim$~200~cMpc.

The phase change due to \heii\ reionization can be probed with absorption in the \heii\ \lymana\ (303.78\AA; hereafter \heii\ \lya) forest, however, very few \heii\ \lya\ forests are observable. Recent work (\citealt{Morrison2019}) showed that the large-scale distribution of high ionization metals such as \ovi\ probed through absorption can compliment and (where necessary) be used as proxy for \heii\ \lya. These measurements showed inhomogeneity on scales from $\sim 10$~cMpc and $\gtrsim 200$~cMpc in the UV background irrespective of the observability of the sources of ionization.

In addition to the study of absorbers associated with structures along a single line-of-sight, there have been attempts to study how a quasar effects gas by studying gas in close lines-of-sight at the same redshift, or the transverse proximity effect. 

A larger emphasis has been placed on the study of the inhomogeneities in the epoch of \hi\ reionization then \heii\ reionization. \citet{Schirber2004} examined 3 quasars from the SDSS Early Data Release with close foreground quasars to measure \hi\ in the \hi\ \lymana\ (1215.67\AA; hereafter \hi\ \lya) forest to determine the intensity of the ionizing background. However they found no evidence for transverse proximity effect in this small sample. Others \citep[e.g.][] {FernandezSoto1995, Crotts1998, Croft2004} conducted similar studies with other lines-of-sight, coming to the same conclusion. \citet{duMasdesBourboux2017} and \citet{Blomqvist2019} utilized the statistical method of cross-correlation of quasars with the \hi\ \lya\ forest flux transmission to study large-scale structure (i.e. the baryon acoustic oscillation). In this context, they fit parameters of the UV fluctuations and transverse proximity effects, assuming a mean free path of 300 $h^{-1}$ Mpc for UV photons. \citet{Goncalves2008} combined high ionization metal line systems within 5 Mpc of quasars with \hi, while determining high quality quasar redshifts and understanding the large-scale environments of the quasars. They suggested that the lack of detections in other studies was most likely the result of inaccurate redshifts or missing environmental information. 

In addition to studying \hi, attempts have been made to study \heii\ proximity zones \citep[e.g][]{Jakobsen2003}. \citet{Worseck2007} used a combination of \hi\ and \heii\ to study the effect of transverse proximity on the shape of the UV background, showing that transverse proximity effect, even when undetectable in \hi, can be detected in \heii\ and metals. \citet{Schmidt2017ApJ} utilized statistical stacking in an effort to detect \heii\ transverse proximity effects. They utilized this statistical detection to place constraints on quasar properties and the mean free path of \heii-ionizing photons.

The active quasar contribution to the UV background can be studied by combining a complete quasar sample with the large-scale 3D distribution of metal species in the IGM, as quasar are expected to dominate large-scales \citep{Furlanetto2008}. However this contribution, as previously mentioned, is dependent on the degree of quasar beaming, light echos \citep{Visbal2008}, and AGN duty cycles \citep{Martini2001}.The effect of quasar duty cycle can be examined using the variations in recombination and relaxation times among various metal ionization species \citep{Oppenheimer2013}. Quasar beaming is likely to add mostly variance to a large blind sample of quasar proximity regions (with perhaps a weak signal associated with small angular separation between the studied gas and the observer). An accurate determination of the impact of the quasar populations promises to provide information on all these effects, with consequences for black hole mass assembly, AGN outflows, and the contribution to the UV background associated with star formation as an alternative source. All these elements are necessary for an understanding of how the universe comes to be ionized and is kept ionized over cosmic time.

%%%%%%%%%%%%%%%%%%%%%%%%%%%%%%%%%%%%%%%%%%%%%%%%%%%%%%%%%%%%
%%%%%%%%%%%%%%%%%%%%%%%%%%%%%%%%%%%%%%%%%%%%%%%%%%%%%%%%%%%%

In this work we explore the direct relationship of quasars and metal absorption/ionization by grouping ISM and CGM absorbers by their proximity to their closest quasars in SDSS-IV/eBOSS using absorber frame stacking. Unidentified quasars bias any trends seen, therefore the identification of a complete quasar sample is critical for the analysis. SDSS-IV/eBOSS was chosen as it is the most complete (75\% to $m_r = 22$ for \lya\ forest quasars) large spectroscopic survey of quasars. The absorbers are identified by \hi\ \lymana\ 1215.67\AA\ in quasar spectra. We build on the absorber frame stacking analysis technique developed by \citet{Pieri2010Stacking,Pieri2014} \citepalias[hereafter][]{Pieri2010Stacking,Pieri2014}. All comoving distances were calculated using Planck 2015 cosmology \citep{Planck2015} with the Astropy cosmology package \citep{astropy:2013,astropy:2018}.

This paper is structured as follows. 
In \Secref{sec:stack_data}, we describe the data and basic treatment of the spectra using masking and continuum fitting. In \Secref{sec:stackMethod}, we outline the stacking methodology and cuts made to the samples. 
\Secref{sec:Pseudo} outlines the creation of a composite spectra using a pseudo-continuum normalization and details the usable portions and spurious signals in the composite spectra.
In \Secref{sec:stack_metals}, we detail the analysis of the metal absorption lines in the composite.
In \Secref{sec:photoion}, we compare our results to the global photoionization rate.
\Secref{sec:stack_Discussion} discusses the results, giving a simple picture.
This is followed by conclusions in \Secref{sec:stack_conclusion}.

\section{The Data}\label{sec:stack_data}

\subsection{Spectroscopic sample building}

Our analysis makes use of the final sample of the eBOSS survey \citep{Dawson2016} taken from Data Release 16 (DR16; \citealt{Ahumada2020}) of the 4th incarnation of the Sloan Digital Sky Survey (SDSS-IV; \citealt{Blanton2017}). This data was taken using the 2.5-metre Sloan telescope \citep{Gunn2006} at the Apache Point Observatory and reduced with the standard SDSS eBOSS reduction pipeline (v5.13.0). The DR16Q sample of quasars \citep{Lyke2020} was compiled through a combination of redshifts and quasar identification procedures including visual inspection, the SDSS eBOSS pipeline, supplementing earlier quasar catalogs (e.g. \citealt{Paris2017}). This sample contains 750,414 quasars, 225,258 of which have $z>2.15$. The locations of all of the quasars are utilized in the analysis, but only the spectra of quasars obtained with the BOSS spectrograph in the eBOSS or SEQUELS footprint ($\sim6250$~deg\textsuperscript{2} of the 10000~deg\textsuperscript{2} BOSS footprint; $\sim6000$~deg\textsuperscript{2} for eBOSS and $\sim250$~deg\textsuperscript{2} for SEQUELS) are studied for gas ionization properties. This reduces the quasars included in the sample to 131,845 with $z>2.15$. This sample was chosen such that the studied gas is in the footprint of higher quasar density compared to the wider BOSS sample and provides a higher completeness. SEQUELS was the BOSS pilot program for eBOSS as part of SDSS-III and therefore also has higher completeness. These spectra have a resolution that varies from R~$\approx$~1650 to R~$\approx$~2500 over a wavelength coverage of 3450-10400 \AA. Calculating for a limiting magnitude of $m_r=22$ (though the sample has a tail of quasars extending out to $m_r = 22.5$ or so) and comparing to the number density given by \cite{Palanque-Delabrouille2016}, our sample is approximately 75\% complete.

Quality cuts are applied to the spectra to ensure that the sample selected is well-understood. Therefore, any spectra with a \lya\ forest signal-to-noise below 0.5 is discarded from the sample. Additionally, absorbers are only selected in regions with a bandwise (100 pixel boxcar smoothed) signal-to-noise $>3$. 
This criterion provides an unbiased sample of absorbers with satisfactory fidelity with respect to the expected true flux transmission selected (see figure 3 of \citetalias{Pieri2014}). While it is critical for a high-fidelity sample of systems to be studied through stacking, we can be more inclusive about associated spectral regions to be stacked. Low signal-to-noise spectra in the stack of spectra are not expected to bias the mean or median stacked spectrum (e.g. artefacts arising due to continuum fitting inaccuracy will not generate sharp absorption-like features).
We reject bands with signal-to-noise $<1$ and $\lambda > 7200$ \AA, where spline fitting cannot cope the combination of low signal-to-noise and the loss of data due to the masking of many sky lines. The region redward of 10200\AA, where the data is typically poor, is also discarded. However, in practise, any spectrum with a signal-to-noise high enough to provide a selected absorber tends to be largely retained, so only a small portion of the data is removed. 

\subsection {Principal component analysis continuum normalisation}
As with all quasar absorption studies, we must normalize the data in order to isolate the absorption signal. Two different normalization methods are used here. We apply the principal component analysis (PCA) of \citet{Lee2012} and \citet{Lee2013} within the \lya\ forest for the selection of the absorbers, updating the DR9Q sample of \citet{Lee2013} for the DR16Q sample. This required masking identified DLAs and the correction of their extended wings. In order to perform this we use the DLA sample provided by the DR16Q catalog of \citet{Lyke2020}. This DLA catalog was produced using the \citet{Parks2018} convolutional neural network architecture with a confidence threshold of 0.9 and $\log \mathrm{N}_\hi > 20.3$. Though we do not use the PCA continuum normalisation outside of the \lya\ forest, the procedure requires good fitting outside of the forest in order to make reliable forest fits. We, therefore, mask strong metal absorption lines outside the \lya\ forest. These masks were created by a 3$\sigma$ outlier flagging of pixels below the continua from spline fitting. While this process does not determine the metal transitions giving rise to the absorption, it is sufficient to identify them for removal prior to PCA fitting. Two sets of eight PCAs, created by \citet{Paris2011} and \citet{Suzuki2005} are fit, initially using an inverse-variance-weighted least-squares fit to flux in $1216\textrm{\AA} < \lambda<1600\textrm{\AA}$ in the quasar rest frame. From the two fits, we select the one showing the lowest reduced $\chi^2$. To fit to the \lya\ forest ($1030\textrm{\AA} < \lambda<1216\textrm{\AA}$ in the quasar rest frame), a mean-flux regulation is applied. The initial PCAs give an estimate of the continuum in the forest but the amplitude is affected by the quasar power-law break and spectrophotometric errors. The PCA continua is modified by 
\begin{equation}
  C_{MF}=C_{PCA}(\lambda_{rest})\times(a+b\lambda_{rest})
\end{equation}
where $a$ and $b$ are the result of a $\chi^2$ fit of the mean flux and the evolution of the mean flux
given by \citet{Faucher-Giguere2008},
\begin{equation}
\langle F\rangle (z) = \exp [-0.001845(1 + z_{abs})^{3.924}].
\end{equation}

\subsection{Spline-based continuum normalisation}
Our PCA continua do not over extend over the entire quasar spectral range of interest (covering only $1030-1600$ \AA). This is satisfactory for the identification of \lya\ absorbers, but associated metals may arise in the full spectral range ($3560-10400$ \AA). Therefore, the spectra used for the stacking step are normalized using an absorption-rejecting cubic spline fit.

The spline fitting is conducted using nodes derived from the median flux in 25\AA\ wide chunks, iteratively rejecting negative deviations from the spline greater than 1~$\sigma$ in the forest and 2~$\sigma$ outside the forest. As spectra are often dominated by noise in the blue, a 100-pixel boxcar smoothing is used, rejecting any smoothed pixels with a smoothed signal-to-noise less then 1. As emission lines are, in general, too sharp to be fit by a smooth spline fit, the peaks of emission features commonly seen in quasar spectra are masked, with the higher and lower bounds given in \autoref{tab:EM_mask} ($\lambda_{Mask}$). We discard a further wavelength range at the edges of the derived spline fitted continuum of scale given by $\pm \lambda_{buffer}$ since spline fits are prone to unreliable deviations at their limits. Finally, we use the same procedure for masking DLAs as described above.

The PCA continua give more reliable and unbiased absolute flux transmission measurements needed for recovery of individual absorbers (such as our sample of \lya\ absorbers to stack). The spline procedure fails to separate large-scale absorption blends from quasar continuum and so can be biased low within the \lya\ forest. On the other hand the stacking procedure below is invariant to such biases since it measures small-scale statistical excess absorption with respect to local regions and such biases are erased in the process of re-normalising the excess absorption with respect to smooth variation in the stacked spectrum (see \autoref{sec:Pseudo}). The spline fitting provides spectra which are smooth with respect to the width of absorbers and remove the bulk quasar shape and amplitude from the resulting spectra and thus avoid unwanted weighting in the stacked spectrum. It also provides additional wavelength coverage, critical to the analysis. This continuum fitting procedure does contribute stochastic uncertainty to the stacking procedure and this is discussed further in section \autoref{sec:stackMethod}.

\begin{table}
  \centering
  \caption{Emission Line Mask Used in the Spline Continua Fitting}
  \label{tab:EM_mask}
  \begin{tabular}{lccc}
    \hline
         Emission Line       &   $\lambda_{rest}$(\AA)& $\lambda_{Mask}$ (\AA)&$\pm \lambda_{buffer}$ (\AA) \\
         \hline
         Ly-$\beta$ &   1033.03         &   1023    -   1041    &   5\\
         Ly-$\alpha$&   1215.67         &   1204    -   1240    &   10\\
         \ion{O}{I} &   1305.42         &   1298    -   1312    &   5\\
         \ion{Si}{IV}&  1396.76         &   1387    -   1407    &   10\\
         \ion{C}{IV}&   1549.06         &   1533    -   1558    &   10\\
         \ion{He}{II}&  1637.84         &   1630    -   1645    &   5\\
         \ion{C}{III}&  1908.73         &   1890    -   1919    &   10\\
         \ion{Mg}{II}&  2798.75         &   2788    -   2811    &   5\\
        \hline
    \end{tabular}

\end{table}

%%%%%%%%%%%%%%%%% %%%%%%%%%%%%%%%%% %%%%%%%%%%%%%%%%% %%%%%%%%%%%%%%%%% 
%%%%%%%%%%%%%%%%% %%%%%%%%%%%%%%%%% %%%%%%%%%%%%%%%%% %%%%%%%%%%%%%%%%% 
%%%%%%%%%%%%%%%%% %%%%%%%%%%%%%%%%% %%%%%%%%%%%%%%%%% %%%%%%%%%%%%%%%%% 
\section{Stacking Methodology}\label{sec:stackMethod}
Following the methodology described in \citetalias{Pieri2014}, absorbers identified in spectra from the DR16 quasar sample are studied through stacking. Entire quasar spectra are stacked based on the position of an absorber of interest (see \autoref{subsec:stacking} below). The entire spectrum is treated as if it is a spectrum of the absorber of interest, even though this is evidently not the case. 
The overwhelming majority of the absorption present in each spectrum stacked is unrelated to the systems selected, but by stacking many such systems this largely uncorrelated absorption only adds a form of noise to the stacked spectrum.
\citetalias{Pieri2014} presented a exploration of their \lya\ selection function and in order to benefit from it here, we used the same data quality requirements (as discussed previously in the previous section). Here we summarise the method used, though a more detailed explanation can be found in \citetalias{Pieri2014}.

\subsection{\texorpdfstring{\lya}{Ly-alpha} absorption system selection}

In order to select absorbers for study from the (PCA normalised) spectra we limit ourselves to the restframe wavelength range 1041\AA\ $<\lambda <$ 1185\AA. The spectra are rebinned every 2 pixels to increase the signal-to-noise and bring the resolution closer to the size of one full width half-maximum (FWHM) element (which is approximately $2.3$ pipeline pixels or 160/kms). In this way we reduce double counting to achieve a reliable bootstrap error analysis. \lya\ forest lines typically have a Doppler parameter of $b\approx 30$\kms \citep[e.g.][]{Hu1995,Rudie2012} or 50\kms FWHM. Since this is significantly less than the spectral resolution the measured flux transmission arises due to a combination of line strength and the blending of lines within the bin. This is made particularly clear where saturated absorption is found despite the fact that single non-damped saturated Lya line cannot generate such a signal in SDSS spectra \citepalias{Pieri2010Stacking,Pieri2014}.
The quasar restframe wavelength range used is the range between the quasar \lya\ and Ly-$\beta$ emission lines and excluding a conservative 7500~km~s\textsuperscript{-1} from the quasar \lya\ emission redshift to avoid the selection of absorbers physically associated with the quasar itself. The blue end limit eliminates the risk of selecting \ovi\ and \lyb\ absorbers. The remaining \lya\ forest is assumed to be the \lya\ absorption caused by intervening gas clouds. While metal absorber contamination occasionally occurs, \citetalias{Pieri2014} analyzed the purity from the stacked spectra and concluded that impurities were of the level of 1-2\% and can be neglected (with some caveats at specified locations in the resulting composite spectrum). The DLA sample provided by the DR16Q catalog of \citet{Lyke2020}, and excluded during our PCA normalisation, is excluded from our absorber identification. This list of absorption systems are separated into various sub-samples in the analysis, as discussed in \autoref{cuts}.

\subsection{Spectral stacking}
\label{subsec:stacking}

Once the sample of absorbers is defined we must rescale wavelength solution of the host spectrum for each marker such that each marker lies at 1215.67\AA. In practise this means dividing the entire spectrum by $(1+z_m)$) where $z_m$ is the redshift of the marker (under the assumption that it does indeed arise due to the \lya\ transition). As explained in \citetalias{Pieri2014}, this de-redshifting procedure requires no interpolation assuming that the entire spectral sample has a fixed $\Delta \log \lambda$ solution.

For the primary analysis, markers with a redshift of $2.4<z<3.1$ are selected, where the minimum redshift is chosen such that associated \ovi\ is always in the observational band (for SDSS spectra). If we did not require this, the systems studied in \civ\ (for example) would not all be observable in \ovi\ and a difference in mean redshift would result. This requirement also ensures that sample size is mostly unchanged irrespective of $\lambda$ over the range of interest (see \autoref{fig:FullStack}). However, for $2.05<z<2.4$ we were not able to provide such sample homogeneity and so we do not study \ovi\ at these redshifts. Here, the minimum redshift selected is the minimum of the \lya\ forest in quasars in BOSS and eBOSS. 

The high and low 3\% outliers in the stack of spectra for every point in the absorber frame grid are clipped. The resulting 2D stack of spectra are then rebinned in the wavelength direction by a factor of 2 pixels (138 km s\textsuperscript{-1}) to reflect the resolution and match the binning of the absorber selection. The median of this resampled grid is then calculated, giving what we call `the stacked spectrum'.
The stacked spectrum has two sources of uncertainty. The first is the pipeline-estimated observational error from the individual spectra. The second arises due to absorption in the stack of spectra not associated with the selected absorber. \citetalias{Pieri2014} found that the error for transitions measured outside the \lya\ forest is of the order of the observational error. However, inside the \lya\ forest the uncertainty is dominated by this contaminating absorption. In order to include this additional source of error here, the stack of spectra is bootstrapped resampled,
where a bootstrap element is one selected absorption system. If a quasar spectrum has multiple absorption systems, then the stack of spectra includes multiple copies of the quasar spectra de-redshifted to the systems's restframe, and are thus treated as separate elements of the bootstrap. \Morrison\ explores the error analysis in more detail and find that 1000 bootstrap realizations are required to completely sample the noise in the stacked spectrum. Each of these bootstrap realisations was treated the same way as the data; they provided a stacked spectrum characterised by the median statistic. The 1-$\sigma$ uncertainty at every wavelength in the stacked spectrum was derived from the standard deviation of this 1000 bootstrap stacked spectra.

Redwards of the \lya\ absorber selected, there is a region that has a mix of spectra with \lya\ forest contamination and without it. To improve sample homogeneity, \citetalias{Pieri2014} produced two different stacked spectra for each sample, one where all \lya\ forest pixels redward of selected \lya\ absorption are not allowed to contribute to the stack of spectra,
and another where they {\it are} allowed to contribute. The choice of which to use was optimised as a function of wavelength based on a comparison of the error estimation.
In this work (see \autoref{fig:FullStack}), the choice is immaterial since the associated metals studied are either always in the \lya\ forest (\ovi\ and \SiIII) or never in the \lya\ forest (\SiIV\ and \civ).

\subsection{Absorption system subsamples}\label{cuts}

The primary interest here is studying the significance of 3D proximity to quasars, but in order to do this we must define consistent and somewhat homogeneous sets of \lya\ absorption systems to study. We focus on two distinct regimes demonstrated in \citetalias{Pieri2014} and further established in \Morrison, a sample of systems showing IGM properties and a sample which is at least partly driven by CGM conditions. The basis for this argument rests on comparison to Lyman break galaxy samples, the presence of low ionization species, which drive one to CGM-like models and the relatively high bias of these systems. 

To explore the properties of only the IGM, only the 2 weakest bands used in \citetalias{Pieri2014} are used, giving a combined flux transmission bin of $0.25\leq$~F(\lya)~$<0.45$. These bands were identified as being predominately associated with IGM absorbers. This is supported by the lack of low ionization features and the presenence of more characteristic high ionization lines. By contrast, the 3 strongest flux transmission bins 
from \citetalias{Pieri2014} ($-0.05\leq$~F(\lya)~$<0.25$) 
were CGM regions with varying purity depending on \lya\ strength, partly due to noise incursions by weaker absorbing pixels that properly belong in the weaker absorption band above.
Therefore, they are described as Mixed CGM stacked spectra for the remainder of the analysis.
The flux transmission bins from \citetalias{Pieri2014}, were combined into 2 single analysis flux transmission bins (IGM and Mixed CGM) to maximize the statistics and facilitate various cuts of the single list of markers

The survey footprint is broken up into HEALPix\footnote{\url{http://healpix.sf.net}} of approximately 1 degree in radius using the HEALpy package with nside=64 \citep{healpy}.
The comoving distance to quasars for every absorber is calculated for quasars within the nearest 8 HEALPix.

The systems are grouped by distance to the closest known quasar. The grouping is set by the distance ranges shown in \autoref{tab:Abs_samp}. The resulting groups of \lya\ absorbers are then utlized to produce stacked spectra. These boundaries were chosen in order that each sub-sample has adequate statistics, while preserving sufficient refinement to probe the shape of the large scale proximity effects. \autoref{tab:Abs_samp} gives the number of absorbers in each sub-sample.

%%%%%%%%%%%%%%%%% %%%%%%%%%%%%%%%%% %%%%%%%%%%%%%%%%% %%%%%%%%%%%%%%%%% 

\begin{table}
    \centering
    \caption{Number of  IGM \& Mixed CGM absorbers used in each sub-sample stack based on comoving distance to nearest quasar}
    \label{tab:Abs_samp}
    \begin{tabular}{lccccccc}
        \hline
        \multicolumn{1}{c}{Separation}  &&\multicolumn{2}{c}{$2.05<z<2.4$} &&\multicolumn{2}{c}{$2.4<z<3.1$}  \\
        \multicolumn{1}{c}{(cMpc)}
                    &&IGM       & CGM   &&IGM    & CGM  \\
        \hline
      All Absorbers &&163003    &72885  &&279765    &137556 \\
        $r\leq30$   &&20532     &10058  &&16161     &8961   \\ 
        $30<r\leq50$&&46431     &20867  &&44920     &22184  \\ 
        $50<r\leq70$&&52520     &23123  &&68570     &33085  \\ 
        $70<r\leq90$&&31523     &13813  &&66706     &32877  \\ 
       $90<r\leq110$&&10025     &4123   &&46036     &22257  \\ 
        $r>110$     &&1972      &827    &&36957     &17972  \\ 
        \hline
    \end{tabular}
\end{table}

%%%%%%%%%%%%%%%%% %%%%%%%%%%%%%%%%% %%%%%%%%%%%%%%%%% %%%%%%%%%%%%%%%%% 

To study the effect of the integrated flux from quasars incident on the absorber, the sample of \lya\ absorbers was alternatively split based on the integrated quasar flux in the SDSS-r band from all neighboring quasars (within 1 degree of the absorber). To facilitate this, effects such as light travel time (i.e. quasar variability) and quasar beaming are neglected, with the basic assumption that luminosity seen on Earth is isotropically radiated and constant with time. The physical distance between the absorber and any given quasar is related to the comoving separation via the cosmological scale factor, $D=a(z)D_c$, calculated at the mean redshift of the pair. Using this, the apparent magnitude of the quasars, assuming isotropically radiating quasars, are calculated at the location of absorber (or marker) with $m_r=5.0(\log_{10}(D)-1)+M_r$. From this, the apparent flux of from each quasar is calculated, $f_r=3631\, \mathrm{Jy}\,\times 10^{-.4m_r}$, and the integrated flux in the SDSS-r band, $F_r=\sum_{qso}f_r$, at each absorber is obtained. The list of markers is then cut into sub-samples of integrated quasar flux evaluated in the SDSS-r band (\autoref{tab:Abs_samp_flux}). These sub-samples of \lya\ absorbers were then used to produce stacks. \autoref{tab:Abs_samp_flux} gives the number of absorbers in each sub-sample. 

%%%%%%%%%%%%%%%%% %%%%%%%%%%%%%%%%% %%%%%%%%%%%%%%%%% %%%%%%%%%%%%%%%%% 

\begin{table}
    \centering
    \caption{Number of IGM absorbers used in each sub-sample composite of integrated r-band flux of all quasars within 1 degree}
    \label{tab:Abs_samp_flux}
    \begin{tabular}{ccc}
        \hline
        \multicolumn{1}{c}{Integrated Flux}  && $2.4<z<3.1$\\
        \multicolumn{1}{c}{(Jy)}        && IGM \\
        \hline
        All Absorbers                   &&279765 \\
        $\leq 361$                      && 20728 \\
        $361 - 458$                     && 29753 \\
        $458 - 475$                     &&  6261 \\
        $475 - 605$                     && 61718 \\
        $605 - 720$                     && 53424 \\
        $720 - 787$                     && 23814 \\
        $787 - 970$                     && 41639\\
        $> 970$                         && 42428\\
        \hline
    \end{tabular}
\end{table}

%%%%%%%%%%%%%%%%% %%%%%%%%%%%%%%%%% %%%%%%%%%%%%%%%%% %%%%%%%%%%%%%%%%% 
%%%%%%%%%%%%%%%%% %%%%%%%%%%%%%%%%% %%%%%%%%%%%%%%%%% %%%%%%%%%%%%%%%%% 
%%%%%%%%%%%%%%%%% %%%%%%%%%%%%%%%%% %%%%%%%%%%%%%%%%% %%%%%%%%%%%%%%%%% 

\section{Creating a Composite via Pseudo-Continuum Normalization}\label{sec:Pseudo}

\begin{figure}
    \centering
    \includegraphics[width=\columnwidth]{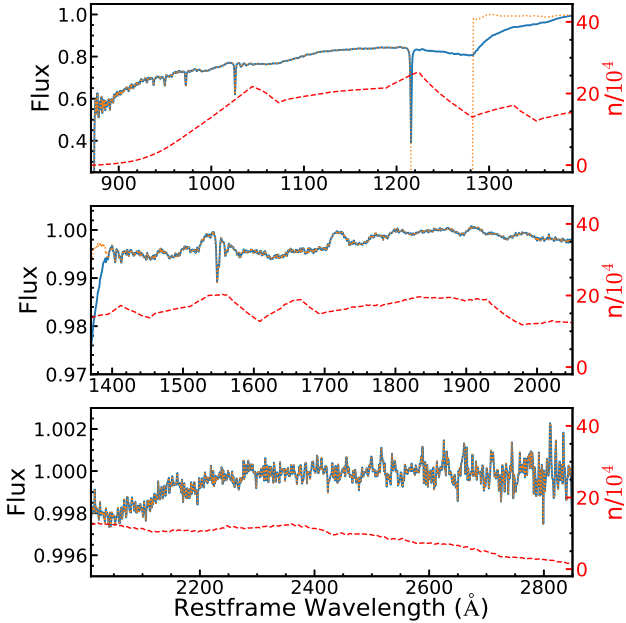}
    \caption{The Median stacked spectra of IGM absorbers with $0.25<F($\lya$)<0.45$ and $2.4 <z<3.1$. The blue line is the stack of all pixels, and the orange dotted line is the stack without \lya\ forest contamination (masking all \lya\ forest pixels redward of 1216\AA\ in the absorber restframe). The dashed-red line is the number of absorbers contribution to the pixels in the unmasked stack. This illustrates the non-unity continuum bluewards of \lya.}
    \label{fig:FullStack}
\end{figure}

The stacked spectra, seen as the blue line in \autoref{fig:FullStack}, clearly shows absorption features associated with the \lya\ absorption systems selected, but it can not yet be regarded as a composite absorption spectrum of these systems. The stacked spectrum of these systems shows various broad absorption effects, that are unrelated to these systems. The most obvious is the broad trough seen from $\sim 900$\AA\ to $\sim 1300$\AA. This arises to the strong contribution of uncorrelated \lya\ forest absorption. Indeed the suppression is at the level of 20-30\%, as approximately expected for mean flux decrement of the \lya\ forest.
In order to study the individual features and arrive at a composite spectrum of them, one has to correct for this for this uncorrelated absorption. We do this by renormalizing the spectrum via a `pseudo-continuum fitting' \citepalias{Pieri2010Stacking}.

The method used by \citetalias{Pieri2014} to create a full pseudo-continuum of the stacks is similar to the process of spline fitting of the individual quasars. However, the stacked spectra is fit with a series of splines for regions of broadly similar broad flux suppression. This series of splines are joined to create a pseudo-continuum. The flux decrements of the calculated pseudo-continuum is added to the stacked spectra, resulting in a composite spectrum where all remaining absorption is assumed to arise due to intrinsic lines. 

The decision to apply the pseudo-continuum normalization in an additive method was chosen as \citetalias{Pieri2010Stacking} argued that, at the resolution of SDSS, contamination absorption would occur within the same resolution element without overlapping the feature of interest in an ideal spectra and thus is more accurately described as additive in flux decrement rather than additive in optical depth. 

There is an error in this process, which is associated with the uncertainty in the stack spectrum and inflates the errors in (pseudo-continuum corrected) composite spectrum. 
\Morrison\ sought to include the uncertainty in this process by performing an `end-to-end' bootstrap analysis in which the realizations were also pseudo-continuum fitted. \Morrison\ found that the additional error due to pseudo-continuum fitting was a small correction. Hence we take the bootstrap errors from the stacked spectra for increased speed in this analysis.

As this study was only focusing on a few features for multiple sub-samples, the process used for the analysis of the metal lines in these stacks deviated from the method of \citetalias{Pieri2014} for metal line analysis, rather then computing a pseudo-continuum of the entire stack, only the features of interest to the analysis were pseudo-normalized. This process is similar to the locally calibrated pixel method, where the flux decrement of a pixel is calibrated by the neighboring pixels being normalized \citep{Pieri2010}. To do this, five rebinned pixels on either side of the absorption feature of interest (\autoref{tab:pseudo_norm}) are used to calculate a linear continuum over these 10 pixels plus the line of interest. This fit was multiplied by the stack, resulting in a composite spectra around the feature of interest. Therefore, the errors calculated from the bootstraps during the stacking are simply propagated through the local continuum fit. 

%%%%%%%%%%%%%%%%% %%%%%%%%%%%%%%%%% %%%%%%%%%%%%%%%%% %%%%%%%%%%%%%%%%% 

\begin{table}
    \centering
    \caption{Wavelength Regions Used to Calculate the Pseudo-Continuum}
    \label{tab:pseudo_norm}
    \begin{tabular}{lccc}
        \hline
          Absorption& & &\\
          Feature &$\lambda_{rest}$(\AA)  & $\lambda_{low}$ (\AA)&$\lambda_{high}$ (\AA) \\
         \hline
         \ion{O}{IV}        &  1031.912             &   1027.7  -   1030    &   1039.5  - 1042      \\
         \ion{C}{IV}        &  1548.202             &   1542    -   1546    &   1553.25 - 1557      \\
         \ion{Si}{IV}       &  1393.76              &   1388.5  -   1391.5  &   1396    - 1399      \\
         \ion{Si}{III}      &  1206.51              &   1202    -   1205    &   1208    - 1211      \\
        \hline
    \end{tabular}
\end{table}

%%%%%%%%%%%%%%%%% %%%%%%%%%%%%%%%%% %%%%%%%%%%%%%%%%% %%%%%%%%%%%%%%%%% 

\subsection{Usable Portions of the Composite and Spurious Signals}
\begin{figure}
    \centering
    \includegraphics[width=\columnwidth]{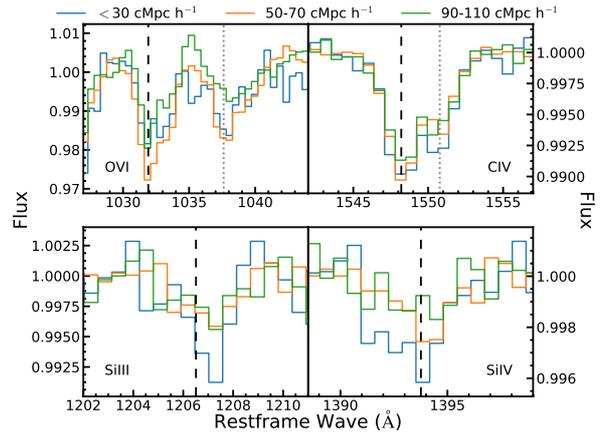}
    \caption{The median composite spectrum for our fiducial case (IGM systems at $2.4 <z<3.1$) focusing on particular metal lines for various quasar proximity with local. The dashed black line is the wavelength of the metal line analyzed, and the gray dotted line (where applicable) is the location of the weaker line of the doublet. }
    \label{fig:IGM_stacks}
\end{figure}

The goal is to probe the effect of quasars on the IGM, hence only the high-ionization metal lines, which are seen in all 5 of the flux transmission bins probed by \citetalias{Pieri2014}, are examined. These are \ion{O}{VI} (1032,1037 \AA\AA), \ion{Si}{IV} (1394,1403 \AA\AA), \ion{C}{IV} (1548,1551 \AA\AA), and \ion{Si}{III} (1206.5 \AA), however in the cases of doublets, only the stronger of the 2 lines is analyzed. \autoref{fig:IGM_stacks} shows the metal feature composite spectra for various quasar proximity scales. Composite stacks for additional scales, and other analysis samples, are included in Appendix \ref{apx:Stacks}. \Morrison\ explores the full population of absorbers in the full composite contributing to the central 2-pixel bin in the composite. However, here the simpler path is taken and we analyze the typical properties of the composite using properties determined by the median composite spectrum.

As previously mentioned, sometimes absorbers due to transitions besides \hi\ \lya\ get selected. This confusion results in spurious features appearing in the composites. Here the wavelength ratio between the misidentified transition and another metal transition is similar to that of \hi\ \lya\ and a metal transition. This can cause confusion and blending. However, our analysis is limited to pixels closer then any potential blending, and therefore is unaffected by such confusion.

In order to characterise the typical metal populations at various physical separation, only the central 2-pixels bins are analyzed to determine the population properties. While one would naively assume that the best way to determine the characteristic properties (flux decrement, column density, and Doppler parameter) of the composite is to do full profile analysis, the full profile is a degenerate mix of contributions such as large-scale structures and metal complexes. 
Therefore, the analysis is conducted using the central 2-pixel bin, which characterizes the typical properties the population, as this is the same scale used for selection of the \lya\ absorbers.

Hereafter composite metal flux transmission means the 2-pixel central pixel bin of the median stack normalized using the method described here.

%%%%%%%%%%%%%%%%% %%%%%%%%%%%%%%%%% %%%%%%%%%%%%%%%%% %%%%%%%%%%%%%%%%% 
%%%%%%%%%%%%%%%%% %%%%%%%%%%%%%%%%% %%%%%%%%%%%%%%%%% %%%%%%%%%%%%%%%%% 
%%%%%%%%%%%%%%%%% %%%%%%%%%%%%%%%%% %%%%%%%%%%%%%%%%% %%%%%%%%%%%%%%%%% 
\section{Metal Absorption and Proximity to Quasars}\label{sec:stack_metals}
Here we present measurements probing the effect of quasar proximity on metal ion column density. We do this by studying the associated metal absorption for sub-sets of absorbers defined by their apparent physical proximity to quasars. We perform this in two ways; initially we split the sample by separation to nearest quasar (and thus assume that the nearest quasar dominates), we then sum the contributions of all quasars to calculate (and then split by) the total integrated r-band flux inferred for each system.

The word `apparent' above is key. Current quasar samples are significantly incomplete and the sample used here is no different despite being the largest survey of \lymana\ forest quasars ever assembled. This means are our ability to group by closest quasar or integrated flux is fundamentally limited. Missed quasars will mean that many systems will have overestimates of the distance to the closest quasar mixed in with systems with correct estimates. On the other hand the impact of missing quasars on the integrated flux will not be so binary leading to varying degrees of lost contribution for essentially all systems. These effects will tend to dilute the signal presented here, and thus a non-detection of sensitivity to quasar proximity is not conclusive, but conversely a statistically significant signal will be robust (even if the signal amplitude is difficult to interpret). Given these limitations, our goal is to demonstrate a method that can be applied to more complete future surveys, and observe early signs of interesting effects.

\subsection{The intergalactic medium system sample}

\begin{figure}
    \centering
    \includegraphics[width=1.\linewidth]{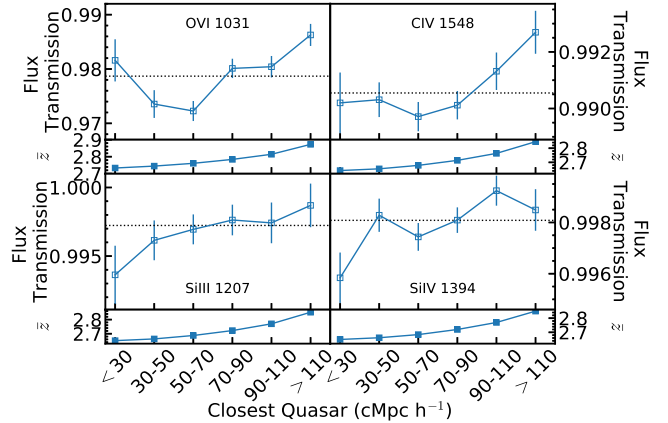}
    \caption{The composite flux transmission of IGM ($2.4 <z<3.1$) metal absorption lines as a function of quasar proximity, with the corresponding mean redshift of the absorbers in each composite spectra.  The gray dotted line shows the composite metal flux transmission of the respective line in the flux transmission of all IGM ($2.4 <z<3.1$) absorbers regardless of quasar proximity.}
    \label{fig:CPA}
\end{figure}

As explained in \autoref{cuts}, we regard our purest sample of intergalactic systems to be defined as those with a \lya\ flux transmission $0.25\leq$~F(\lya)~$<0.45$ based on the results of \citetalias{Pieri2014}. This result is based on an investigation of the selection function based on comparisons to high quality spectra, samples of Lyman break galaxies, and simulations in the redshift range $2.4<z<3.1$. We therefore regard this range in flux transmission and redshift to provide our fiducial results. \autoref{fig:CPA} shows the composite metal flux transmission for this fiducial sample ($2.4<z<3.1$) in sub-samples as a function of separation to closest quasar. As shown in \autoref{tab:chisq}, \ovi\ shows a greater than 4$\sigma$ detection of dependence on quasar proximity, \civ\ and \SiIV\ are both significant at approximately the 2$\sigma$ level, while \SiIII\ is consistent with the null.

The \ovi~1031 transition follows a trend of stronger absorption for closer quasar proximity for scales $\gtrsim 50$ cMpc h$^{-1}$. Assuming that closer quasar proximity equates to additional ionizing photons, this would appear to indicate a greater degree of ionization from \ion{O}{V} to \ovi. On the other hand, for $\lesssim 50$ cMpc h$^{-1}$ there are tentative signs of weakening \ovi\ absorption to the smallest scales studied, perhaps indicating yet further ionization to \ion{O}{VII} dominates.

\civ\ and \SiIV\ shows weak trend of stronger absorption for closer proximity. Again assuming that the closest quasars dominate overall and that their closeness result in a significant increase in ionizing photon intensity for the studied systems, some initial interpretation is possible. These results suggest a greater degree of ionization from \ion{C}{III} to \civ\ and from \SiIII\ to \SiIV, though some balancing may be necessary with regards to silicon given the lack of a significant opposite dependence for \SiIII\ perhaps meaning that ionization from \ion{Si}{II} to \SiIII\ is also needed.

For each sub-sample in each panel of \autoref{fig:CPA}, we provide the mean sample redshift. These curves are the same for every panel except for \ovi, because some systems in each sub-sample have redshifts too low for measurement of \ovi\ absorption. However, we show one curve per measurement panel for simplicity. Overall, our results shows a slight trend towards higher redshifts for larger separations. This arises due to the fact that there is a higher density of quasars per unit comoving volume at lower redshift. Overall, the redshift range of quasars used is sufficiently limited and the resulting difference in mean absorber redshift is sufficiently small that redshift evolution is expected to be weak (\citealt{Schaye2003, Aguirre2004, Aguirre2008}).

\begin{figure}
    \centering
    \includegraphics[width=1.\linewidth]{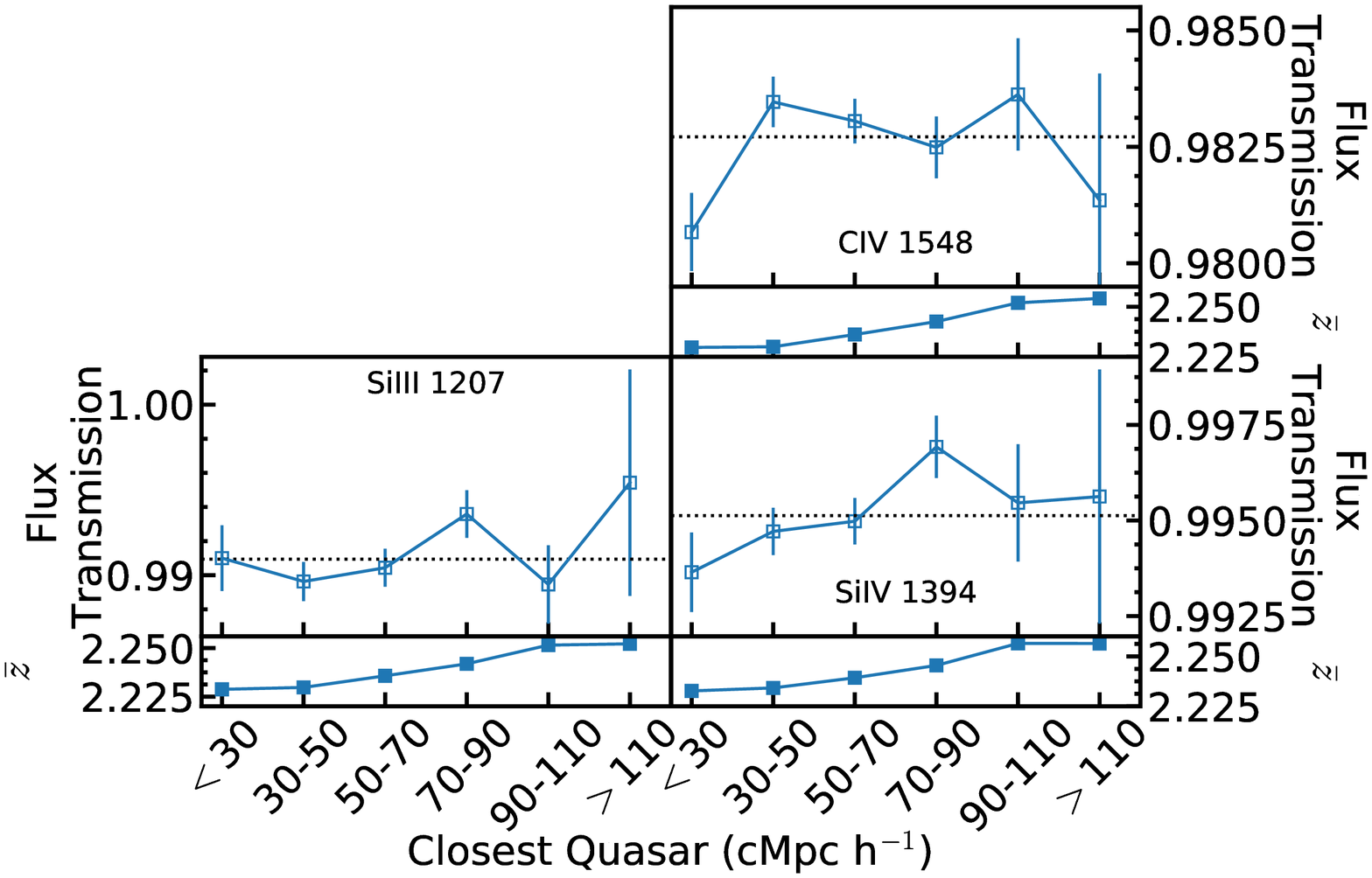}
    \caption{Same as \autoref{fig:CPA} but for low-$z$ ($2.05<z<2.4$) IGM metal flux absorption lines.}
%    \caption{The composite flux transmission of low-$z$ ($2.05<z<2.4$) IGM  metal flux absorption lines as a function of quasar proximity, with the corresponding mean redshift of the absorbers in each composite spectra. The gray dotted line shows the composite metal flux transmission of the respective line in the composite of all low-$z$ IGM absorbers regardless of quasar proximity. }
    \label{fig:CPA_lz}
\end{figure}

\autoref{fig:CPA_lz} shows the composite flux transmission of \civ, \SiIII, and \SiIV\ for the low redshift IGM sample ($2.05<z<2.4$). None of the species studied here show sensitivity to quasar proximity at the 2$\sigma$ level. This is largely because the most significant case, that of \ovi, is inaccessible at these redshifts. It is also because the already weak signal for \civ\ and \SiIV\ in our higher redshift fiducial sample is weaker still here and no longer reaches this threshold. Signs of a trend towards stronger \civ\ absorption when a quasar is closer in particular is no longer apparent with deviations from the mean absorption showing no signs of any trend.

%%%%%%%%IGM Main z

%%%%%%%%IGM low z

\begin{table*}
    \centering
    \caption{The $\chi^2$ test for agreement with the null case where the sub-samples split by quasar proximity are statistically consistent with mean for the full sample. There are 5 degrees of freedom (accounting for the loss of one degree in comparing with the mean. Note that 1$\sigma$ is 5.9, 2$\sigma$ is 11.3 and 3$\sigma$ is 18.2, 4$\sigma$ is 26.8, 5$\sigma$ is 37.1.}
    \begin{tabular}{lccccc}
        \hline
         Ion            & IGM  High-$z$ & IGM       & Mixed CGM   & Mixed CGM   & IGM\\
                        &   (Fiducial)   &low-$z$    &  High-$z$     & low-$z$   & Integrated Intensity\\
         \hline
         \ovi\ 1031     & 32.1  & $\cdots$  & 43.4      & $\cdots$  &17.6\\
         \civ\ 1548     & 12.9  & 9.3       & 33.6      & 9.3       &7.2\\
         \SiIII\ 1207   & 4.5   & 6.0       & 9.1       & 18.8      &3.8\\
         \SiIV\ 1394    & 10.9  & 7.5       & 5.1       & 5.4       &15.7\\
         \hline
    \end{tabular}
    \label{tab:chisq}
\end{table*}

\subsection{The Mixed CGM system sample}
While the goal is to study the impact of large-scale proximity to quasars using IGM metal absorption, the mixed CGM sample identified in \citetalias{Pieri2014} and selected with $-0.05\leq$~F(\lya)~$<0.25$
may also prove useful. As previously noted, a significant contribution to the high ionization metal lines in the CGM bins is still expected to be from the IGM. In some cases these systems are expected to be IGM absorbers proximate to galaxies and therefore may be dominated by galaxies as a source of ionization and insensitive to quasar proximity. In other cases they are IGM contaminants to the CGM selection. On the one hand the definition of this sample is more complex, on the other the strength of the absorption may boost the measurement accuracy.

\begin{figure}
    \centering
    \includegraphics[width=1.\linewidth]{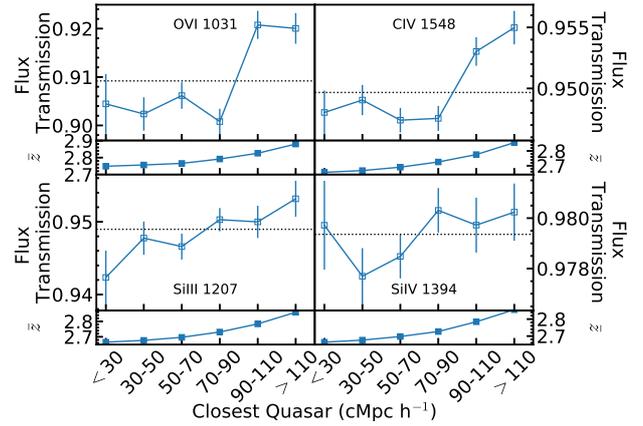}
    \caption{Same as \autoref{fig:CPA} but for Mixed CGM ($2.4 <z<3.1$) metal flux absorption lines.}
    %    \caption{The composite Mixed CGM ($2.4 <z<3.1$) metal flux transmission as a function of quasar proximity, with the corresponding mean redshift of the absorbers in each composite spectra. The gray dotted line shows the composite metal flux transmission of the respective line in the composite of all CGM ($2.4 <z<3.1$) absorbers regardless of quasar proximity.}
    \label{fig:CPA_CGM}
\end{figure}

\begin{figure}
    \centering
    \includegraphics[width=1.\linewidth]{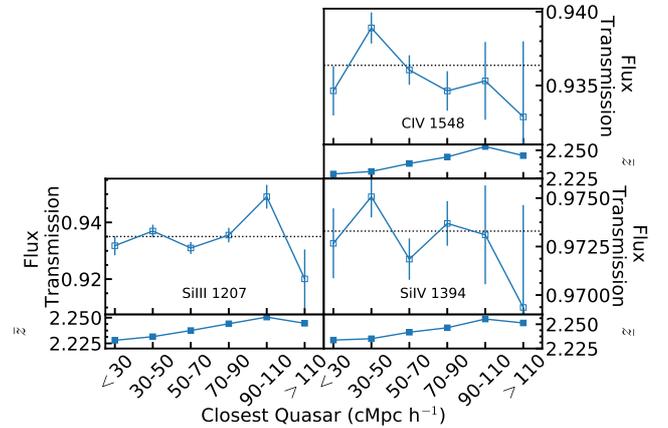}
    \caption{Same as \autoref{fig:CPA} but for low-$z$ ($2.05<z<2.4$) Mixed CGM metal flux absorption lines.}
    %\caption{The composite low-$z$ ($2.05<z<2.4$) Mixed CGM metal flux transmission as a function of quasar proximity, with the corresponding mean redshift of the absorbers in each composite spectra. The gray dotted line shows the composite metal flux transmission of the respective line in the composite of all low-$z$ CGM absorbers regardless of quasar proximity.}
    \label{fig:CPA_lz_CGM}
\end{figure}

\autoref{fig:CPA_CGM} is the Mixed CGM equivalent plot to our fiducial result and \autoref{fig:CPA_lz_CGM} is the Mixed CGM equivalent to the low redshift plot.

At higher redshifts (\autoref{fig:CPA_CGM}) clear signal is seen for stronger absorption seen with closer proximity for \ovi, with a step-like transition at $\sim 90$ cMpc h$^{-1}$ here. In contrast to the fiducial IGM-only result, there are no signs of a turnover at $\sim 50$ cMpc h$^{-1}$, however, the \ovi\ absorption signal does seem to plateau on scales below $\sim 80$ cMpc h$^{-1}$. Overall \ovi\ absorption shows a greater than 5$\sigma$ detection of sensitivity to quasar proximity here. \civ\ seems to follow a similar pattern to the fiducial case, though the significance of the detection of quasar proximity is higher (greater than 4$\sigma$). Hints of a signal in \SiIII\ also follow a similar trend though they approach 2$\sigma$ here. \SiIV\ show now significant dependence on quasar proximity here.

As before, there is a weak and insignificant drift in mean redshift as a function of distance to nearest quasar and \ovi\ mean redshifts are on the whole a little higher than for the other species but none of this approaches the level that might indicate that they could be a cause for the signal seen.

\autoref{fig:CPA_lz_CGM} shows a hint of signal of a reversal for \civ\ with respect to the other measurements whereby \civ\ absorption may be weakening where quasars are more proximate. \SiIII\ appears to peak at around $\sim 100$ cMpc h$^{-1}$ with a 3$\sigma$ significance. \SiIV\ shows no significant signs of a quasar proximity dependence.

%%%%%%%%CGM Main z

%%%%%%%%CGM low z

\subsection{Integrated Effect of Proximate Quasars}
\label{subsec:results_integ}

Since every absorber experiences different ionizing radiation depending on the number of proximate quasars, and their respective distance, an attempt was made to explore the relationship between the integrated flux of quasar radiation incident on the absorber and metal absorption. For this, we produced a set of composite spectra with absorbers sub-sampled based on the integrated flux from proximate quasars at the location of the absorbers for the fiducial redshift range and IGM absorber sample. For simplicity, we calculated the integrated flux in the SDSS-r band, 
\begin{equation}
    F_r=\sum_{qso}f_r 
    = \sum_{qso} 3631\, \mathrm{Jy}\,\times 10^{-.4m_r},
\end{equation}
for each absorber, where $m_r$ is the apparent quasar magnitude seen by the absorber (See \autoref{sec:stack_data} for more details). The same sample splits of \lya\ absorption strength and redshift employed in the fiducial analysis are utilized here.

We assumed that r-band flux observed was radiated isotropically and integrated all the contributions based on the distance to the system selected. In this, we ignored that this flux is not the ionizing flux, and that the energy is biased with respect to the required energy. The true energy is both redshift and species dependent.

\begin{figure}
    \centering
    \includegraphics[width=1.\linewidth]{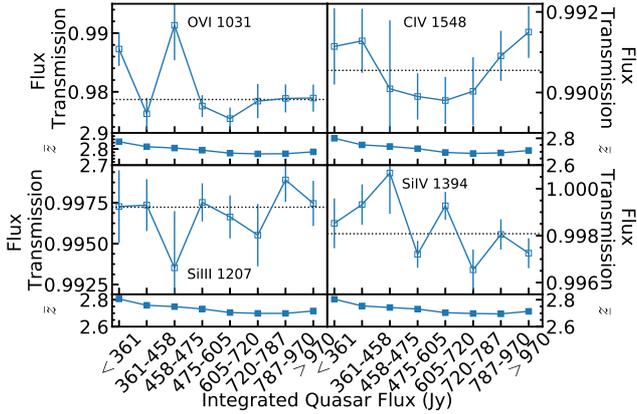}
    \caption{
    The IGM ($2.4 <z<3.1$) metal flux transmission as a function of integrated quasar flux of all quasars with in 1 degree on the sky, with the corresponding mean redshift of the absorbers in each composite. This is an alternative estimator of quasar proximity to that shown in \autoref{fig:CPA}. In comparing the two it should be noted that the sign of the expected effect is inverted in the x-axis since a smaller distance to quasars will tend to lead to a larger flux. The gray dotted line shows the stacked metal flux transmission of the respective line in the composite spectra of all IGM ($2.4 <z<3.1$) absorbers regardless of quasar proximity.}
    \label{fig:CPA_int}
\end{figure}

\autoref{fig:CPA_int} shows the composite metal flux transmission for the fiducial IGM sample $(2.4<z<3.1)$, as well as the mean redshift for each composite with sub-samples of integrated r-band flux. It broadly mirrors what is seen with the `distance to closes quasar' approach shown in \autoref{fig:CPA}, but appears to do so with poorer statistics. It should be noted, of course, that when comparing `distance to closes quasar' and integrated r-band flux, the expected signal is inverted. As before, \ovi\ absorption appears to be stronger when quasar proximity is more pronounced and the disagreement with the null result is at the $\sim 3\sigma$ level here. Again \SiIV\ absorption sensitivity to quasar proximity is somewhat significant (this time to greater than 2$\sigma$ significance), but \civ\ does not show a significant signal despite a mildly significant detected seen for the fiducial estimator. \SiIII\ continues to show no significant effect.

It would appear that this estimator of quasar proximity is more sensitive to incompleteness of the quasar sample and therefore dilutes the signal present in the data more severely. Again differential redshift drift between the sub-samples is minimal.

%%%%%%%% Intensity

\section{Comparison to the Global Photoionization Rate}
\label{sec:photoion}

Here we set out a useful straw-man model to explore our results. Let us assume a simple scenario were all quasars contribute the same ionizing flux, that where inhomogeneity exists it is dominated by one local quasar and the associated metal species abundance is entirely homogeneous. In such a scenario, the mean metal absorption seen in our analyses should be reached at the scale where the local quasar contribution to the photoionization rate reaches that of the UV background. We build such a simple model and compare to our results.

We calculate the photoionization rate of various ionization species of oxygen, carbon and silicon as a function of quasar distance using the luminosity of an average quasar \citep{Lusso2015}, a power law spectral slope determined from quasar stacking \citep{Lusso2015,Shull2012,Stevans2014,Scott2004} and the photoionization cross-section \citep{Verner1996}. We compare this to the photoionization from the UV background of quasars and galaxies \citep{HM2012}. 

\begin{figure}
    \centering
    \includegraphics[width=1.\columnwidth]{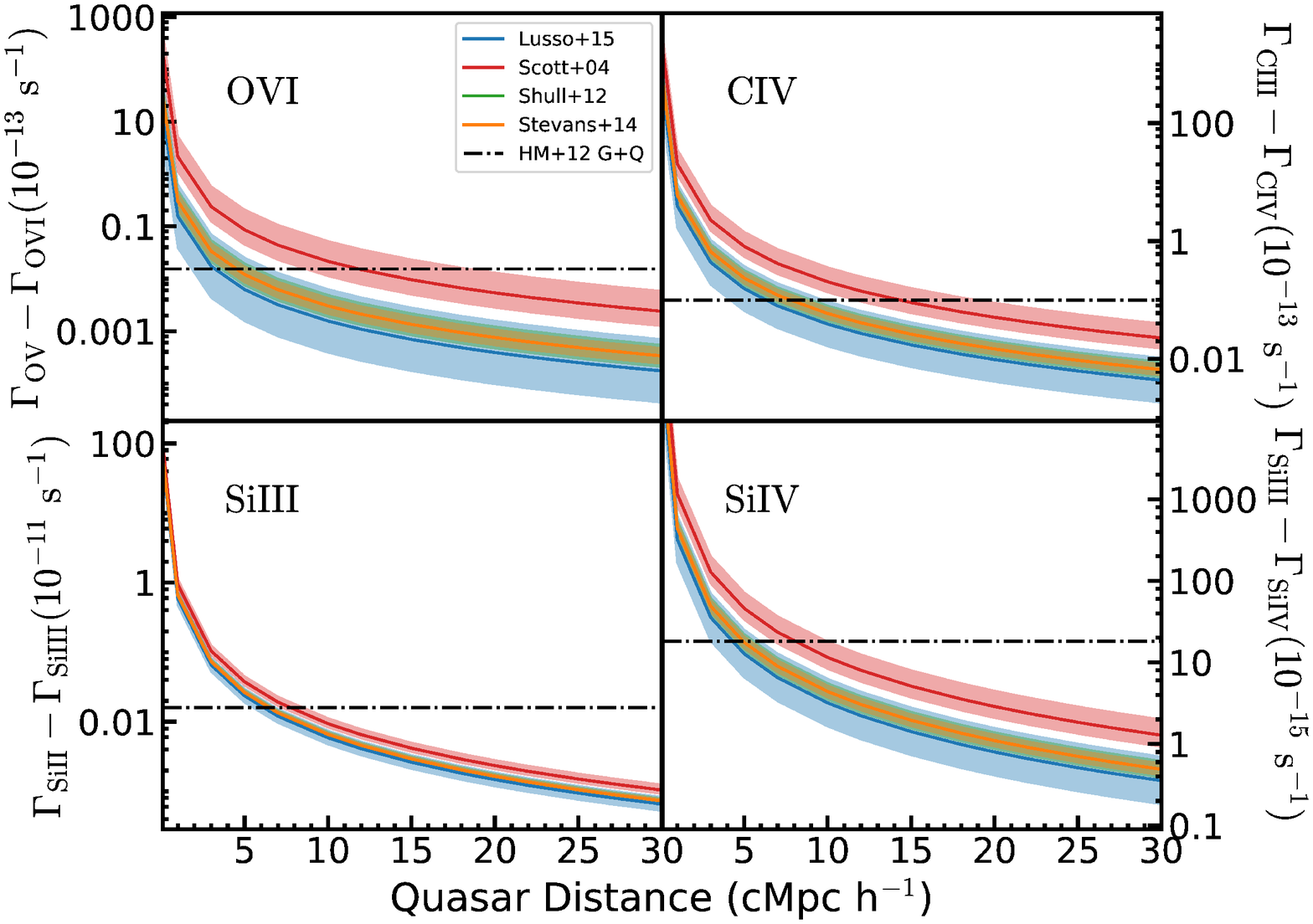}
    \caption{The net photoionization rate ($\Gamma_\mathrm{in}-\Gamma_\mathrm{out}$) of \ovi, \civ, \SiIII, \& \SiIV\ due to a quasar as a function of distance using various quasar power law slopes from \citet{Lusso2015} (blue), \citet{Scott2004} (red), \citet{Shull2012} (green), and \citet{Stevans2014} (orange). This assumes a average quasar flux of $F(1122.0\text{\AA})=1.8\times 10^{-15}$ and a photoionzation cross-section derived from \citep{Verner1996}. The dot-dashed black line is the photoionzation rate from the galaxy+quasar UV background from \citep{HM2012} evaluated at $z=2.829$. 
    }
    \label{fig:photoionNet}
\end{figure}

\autoref{fig:photoionNet} illustrates the effective net photoionization rate, compared to the UV background, for \ovi, \civ, \SiIII, and \SiIV. It illustrates that close to quasars there can be an excess in the photoionziation rate for these metals, while at large distances a decrease compared to the average UV background due to a dearth of close quasars. However, as quasars spectra are characterised by a range of luminosities in addition to a range of power laws, the excess close to quasars could be greater than is illustrated here. 

At close distances to quasars there exists an excess in the photoionization rate of all of the ions analyzed here. However, the scales projected by the model for the transition from an excess to a deficit in metal species abundance ($\sim 10$~cMpc h$^{-1}$) is inconsistent with the scale of the same transition in our results ($\sim 90$~cMpc h$^{-1}$). While our distance to the closest quasar might be statistically overestimated, it is implausible to overestimate to the degree required to agree with these models and still see a signal. The fact that we see a signal on scales incompatible with this model, suggest that this model is too simple. A potential bias, neglected in this simple picture, arises from the fact that the  photoionization rates are calculated using the average quasar luminosity, therefore they are underestimated for the brighter quasars. This goes to show that the quasar luminosity function has a major impact, and that the variations appear to be driven by a small population of bright quasars, perhaps combined with post-helium reionization.

Note that the results of \autoref{subsec:results_integ} are not directly comparable to the UV background models discussed here, and shown as the dot-dashed black line in \autoref{fig:photoionNet}, as those model values are based on the integrated intensity of the photons directly associated with the transition and not the SDSS-r band intensity.

%%%%%%%%%%%%%%%%% %%%%%%%%%%%%%%%%% %%%%%%%%%%%%%%%%% %%%%%%%%%%%%%%%%% 
%%%%%%%%%%%%%%%%% %%%%%%%%%%%%%%%%% %%%%%%%%%%%%%%%%% %%%%%%%%%%%%%%%%% 
%%%%%%%%%%%%%%%%% %%%%%%%%%%%%%%%%% %%%%%%%%%%%%%%%%% %%%%%%%%%%%%%%%%% 
\section{Discussion}\label{sec:stack_Discussion}

The primary goal of this work was to measure large-scale UV background inhomogeneity in three dimensions. This was attempted through stacking of \lya\ forest absorbers measuring the associated variation in the metal absorption strength with respect to apparent proximity to quasars. In light of the limitations of current massive spectroscopic surveys of quasars this study is chiefly a demonstration that promises to be particularly impactful in upcoming generation of massive spectroscopic surveys. Nevertheless this study allows some early conclusions to be drawn.
Clearly significant effects are seen on scales of $\sim 30$ cMpc h$^{-1}$ to $\sim 100$ cMpc h$^{-1}$ even though this is much larger than one might naively expect based on simple straw-man models such as that presented in \autoref{sec:photoion}. Overall our results show either a non-detection of sensitivity to quasar proximity or increased high ionization species absorption closer to quasars. However, it is unclear whether this arises due to more intense UV radiation near quasars or a hardening of the the UV radiation spectrum (or indeed a combination of the two).

In this section we will focus initially on our fiducial IGM, $0.25\leq$~F(\lya)~$<0.45$ and $2.4<z<3.1$, and then compare this to mixed CGM samples, an alternative approach of assessing the impact of quasar proximity through the calculation of integrated incident flux. We will also discuss the limitations of this work and how we hope to address them in future surveys.

The sample of quasars used here is only approximately 75\% complete to a limiting magnitude of $m_r = 22$. Beyond this limit there is a tail to $m_r = 22.5$ and even $m_r = 23$ (where the eBOSS completeness drops off dramatically). As a consequence these results have large potential biases due to quasar incompleteness. For each absorber included, it is possible that there is a closer unidentified quasar than the one used to define the quasar proximity sub-sample to which it belongs. Therefore, the larger separation composites will most likely be contaminated by closer separation systems. This will also reduce the statistics in the closer separation bins, reducing the significance in those bins. Hence our fiducial result systematically over estimate the distances and blur the differential measurement we seek. For this reason, only preliminary interpretation will be explored here and limited to the signal detection and difference rather than amplitude.

The composite metal flux transmission for the fiducial IGM analysis ($0.25\leq$~F(\lya)~$<0.45$ and $2.4<z<3.1$) is broadly consistent with the trend that closer quasars result in more ionizing photons and therefore more ionzation for high ions \ovi, \civ, and \SiIV. The exception is that on the smallest scales studied \ovi\ appears to invert this trend and shows weaker absorption for $\lesssim 50$~cMpc h$^{-1}$ sub-sample. This could be explained \ovi\ being ionized further to \ion{O}{VII}. This is an additional effect not seen in the simple model of \autoref{sec:photoion}, which if confirmed would reinforce the picture that rare bright sources play a key role on the scales that we begin to probe here (perhaps combined with incomplete helium reionization effects).

Our most statistically significant detection of quasar proximity effects arise in what we call the Mixed CGM sample (greater than 5$\sigma$ for \ovi\ and greater than 4$\sigma$ for \civ). They are so called because when these stronger absorbers are found in the moderate resolution SDSS spectra they typically arise in regions within or near the CGM of Lyman break galaxies and their extended absorption analogues. Since the \lya\ signal selected is stronger, it is natural that the metal signal is stronger; the mean signal in these samples is between 4 and 10 times stronger than the fiducial results. As a result, although the quasar proximity signal is stronger here, one cannot claim that quasars have stronger ionizing impact on extended CGM regions based on this work alone. Further investigation is needed to ascertain the comparative impact on IGM and CGM regions. It is, however, striking that a clear signal is seen in both \ovi\ ($> 5\sigma$) and \civ\ ($> 4\sigma$) at $\sim 90$ cMpc h$^{-1}$. The \civ\ effect is similar to that seen for the IGM fiducial sample. \ovi, on the other hand, appears to show a different trend, and the turnover in \ovi\ absorption strength at $\sim 50$ cMpc h$^{-1}$ is not apparent for the Mixed CGM sample. The weak \SiIV\ signal of the IGM fiducial sample is not apparent here but again this may be statistical in origin since the weak \SiIV\ in the fiducial may have been spurious.

On the face of it, the low redshift version of both the fiducial IGM and the Mixed CGM samples appear to show little more than noise. By comparing both with the higher redshift counterparts, however, hints of a pattern may be emerging. Clearly we must limit ourselves to discussion of \civ, \SiIII, and \SiIV\ since \ovi\ is inaccessible at these lower redshifts. For both the IGM and Mixed CGM samples, a trend in \civ\ at high redshift becomes insignificant at low redshift. The change is weak for the IGM sample, but is very pronounced for the Mixed CGM sample (going from $> 4\sigma$ to $< 2\sigma$). Our high redshift sample corresponds to the tail of helium reionization and an epoch where residual UV background inhomogeneities (particularly for hard photons) are still expected on large scales. Our low redshift sample appears to probe the Universe significantly after the end of \heii\ reionziation where the mean free paths of the UV photons is once again much larger than the separation of sources. Our results hint at this trend since ionization from \ion{C}{III} to \civ\ is sensitive to hard photons and the trend of significant weakening of \civ\ signal maybe be consistent with declining UV background inhomegeneity as the universe leaves the post-helium reionization epoch \citep{HM1996,Agofonova2007}. Also there may be signs of \SiIII\ signal beginning to emerge at low redshifts, again particularly for the Mixed CGM case (going from $< 2\sigma$ to $> 3\sigma$). The comparison of \civ\ and \SiIII\ may suggest that this evolution should not be quickly dismissed as simply a decline in statistics at low redshift. Neither is it conclusive, however, and these tentative signs of evolution are chiefly an indication of what can be explored fruitfully in future data.

One must keep in mind a couple of caveats when comparing low and high redshifts. Firstly, the classification of blended absorbers as IGM or CGM by \citetalias{Pieri2014} was only studied for our high redshift sample, and treating the \lya\ selection function for the lower redshift sample is a somewhat untested extrapolation (one motivation, the presence of low ionisation species in optically thin gas, is seen for this low redshift sample). Furthermore, the quasar sample used here is a magnitude limited sample, but it is preferable to use a luminosity limited sample. The size of the analysis bins, in redshift space, minimizes this effect, however, the fiducial redshift sample ($2.4<z<3.1$) is large enough that there is a significant difference in the luminosity distance correction between quasars at the low and high end of the redshift bin. The low redshift bin is not as susceptible to this unwanted dependence as the difference in correction is not as strong. This dependence will also contribute to the differences between the high and low redshift samples, causing confusion mismatches that preclude a detailed comparison at this stage.

The use of integrated quasar flux shows results consistent with those using distance to closest observed quasar (noting the inverted x-axis for the direction of increased ionizing flux).
They are, however, noisier, indicating that they suffer more from the impact of incompleteness in the sample. 

We have assumed throughout this work that the noise in the analysis is Gaussian. It should be noted that when a study performs many tests for signal in noisy data, like ours, spurious effects become more probable (a form of `look elsewhere effect'. In our 18 tests, approximately one $2\sigma$ apparent detection is expected due to noise alone, assuming they are independent. Of course our 4 integrated flux results are not independent of our fiducial analysis, given their common sample, but the fact remains that our weak $2\sigma$ detections should be treated as very preliminary. On the other hand three of our $>3\sigma$ detections are not marginal and our \ovi\ results are particularly robust. The overall the picture presented in \cite{Morrison2019} of inhomogeneities in the \ovi\ on several 10s of cMpc is strongly supported here, despite the very different data and analysis methods used. Indeed, we are able to go further than that work and confirm that these inhomogeneities are at least correlated with quasar positions.

The expanded sample of quasars provided by WEAVE-QSO \citep{Pieri2016} will enable the sub-samples based on quasar proximity to probe smaller scales. To compare the amplitudes of the \ovi\ signals would require large-scale radiative transfer simulations with a prescription for outflows and realistic UV inhomogeneities to address the potential pre-enrichment of the IGM. Within the J-PAS \citep[Javalambre-Physics of the Accelerated Universe Astrophysical Survey;][]{JPAS} footprint, and even more so within the HETDEX \citep[Hobby-Eberly Telescope Dark Energy Experiment][]{Adams2011} spring field, in WEAVE-QSO, the identified quasar populations, whether observed with WEAVE or just identified with J-PAS, will facilitate the addition of a luminousity cut, eliminating the unwanted redshift dependence currently contaminating the SDSS-IV/eBOSS analysis.

%%%%%%%%%%%%%%%%% %%%%%%%%%%%%%%%%% %%%%%%%%%%%%%%%%% %%%%%%%%%%%%%%%%% 
%%%%%%%%%%%%%%%%% %%%%%%%%%%%%%%%%% %%%%%%%%%%%%%%%%% %%%%%%%%%%%%%%%%% 
%%%%%%%%%%%%%%%%% %%%%%%%%%%%%%%%%% %%%%%%%%%%%%%%%%% %%%%%%%%%%%%%%%%% 
\section{Conclusion}\label{sec:stack_conclusion}

Large-scale modulation of the UV background and its dependence on quasars was explored through study of the homogeneity of metal ionization species. This was achieved through absorber frame stacking of \lya\ systems in SDSS-IV/eBOSS with varying three-dimensional proximity to quasars. The goal was to study differences in the associated metal absorption in the resulting composite spectra. We do this for two classes of \lya\ absorption systems: intergalactic medium absorbers and those which have revealed strong CGM properties (which we call a `Mixed CGM sample'). We also perform the analysis in two redshift windows: a high redshift sample at $2.4<z<3.1$ and a low redshift one at $2.05<z<2.4$. Only our high redshift window provides data for our most informative species, \ovi, and this, combined with its greater homogeneity, leads us to characterise the high redshift IGM sample as fiducial for this work. We primarily characterise quasar proximity through the distance to closest quasar but we also attempt to do this by calculating the integrated flux from quasars incident on our \lya\ absorber sample.

The limited completeness of the eBOSS quasar sample ($\sim 75\%$) makes detailed interpretation challenging but some conclusions can be drawn even in this stage as follows:
\begin{enumerate}
  \item Strong inhomogeneity is seen is the metal species studied as quantified by the null $\chi^2$ test of agreement with the scale-free mean metal absorption.
  \item The species \ovi\ shows the strongest effect with $>4\sigma$ detections of structure in both the fiducial IGM sample and the CGM sample for the apparent distance to closest quasar. Broadly speaking, smaller scales give rise stronger \ovi, though at $\lesssim 50$~cMpc h$^{-1}$ there is an apparent reversal in the trend such that less \ovi\ is seen.
  \item \civ\ also shows a detection of increased absorption closer to quasars at high redshift, particular in the CGM case ($>4\sigma$), while \SiIV\ and \SiIII\ show only weak dependence.
  \item At low redshift, the only significant detection of dependence on quasar proximity is \SiIII\ ($>3\sigma$) in the CGM sample.
  \item There are there are tentative signs of weaker sensitivity to quasar distance at low redshift, which is consistent with an increase in mean free paths of UV photons between $z\sim 3.1$ and $z\sim 2.05$, large enough that such structures are erased.
  \item Probing quasar proximity by integrated quasar r-band flux provides consistent but statistically weaker results by comparison to distance to closest quasar.
  \item The results support the findings of \cite{Morrison2019} that \ovi\ shows inhomogeneity on scales of several 10s of comoving megaparsecs, reaching even $\gtrsim 100$~cMpc h$^{-1}$. This work goes a step further, however, confirming expected connection to the quasar population.
\end{enumerate}

Overall, we find that using these techniques to explore the large scale effects of quasars on the UV background is promising, with trends broadly consistent with what one would naively expect. Our straw-man model is not consistent with our results, indicating that real modelling is needed to draw detailed quantitative conclusions. Given the incompleteness in the quasar sample used, more sophisticated modelling would have to include this incompleteness and the bias that results. However, next generation surveys in the process of beginning, will address this survey limitation and render quantitative conclusions more accessible. J-PAS \citep[Javalambre-Physics of the Accelerated Universe Astrophysical Survey;][]{JPAS} will identify complete samples, which will be followed up spectroscopically by WEAVE-QSO (\citealt{Pieri2016}) as part of the wider WEAVE survey \citep[][Jin et al. in preparation]{Dalton2012}
%,Dalton2014,Dalton2016a,Dalton2016b} 
to magnitudes as faint as $r\sim 23.5$. Dark Energy Spectroscopic Instrument (DESI; \citealt{DESI2016}) will provide additional spectra. In light of these incoming surveys, the methods presented here will be performed to higher fidelity and allow a wider range of tests to be performed exploring the impact of quasar luminosity and three-dimensional geometry.

%%%%%%%%%%%%%%%%%%%%%%%%%%%%%%%%%%%%%%%%%%%%%%%%%%
\section*{Acknowledgements}

We thank Joop Schaye, Jamie Bolton, and Clotilde Laigle for useful comments and suggestions. 
SM and MP were supported by the A*MIDEX project (ANR-11-IDEX-0001-02) funded by the ``Investissements d'Avenir'' French Government program, managed by the French National Research Agency (ANR), and by ANR under contract ANR-14-ACHN-0021.

Funding for the Sloan Digital Sky Survey IV has been provided by the Alfred P. Sloan Foundation, the U.S. Department of Energy Office of Science, and the Participating Institutions. SDSS-IV acknowledges support and resources from the Center for High-Performance Computing at the University of Utah. The SDSS web site is www.sdss.org.

SDSS-IV is managed by the Astrophysical Research Consortium for the Participating Institutions of the SDSS Collaboration including the Brazilian Participation Group, the Carnegie Institution for Science, Carnegie Mellon University, the Chilean Participation Group, the French Participation Group, Harvard-Smithsonian Center for Astrophysics, Instituto de Astrof\'isica de Canarias, The Johns Hopkins University, Kavli Institute for the Physics and Mathematics of the Universe (IPMU) / University of Tokyo, the Korean Participation Group, Lawrence Berkeley National Laboratory, Leibniz Institut f\"ur Astrophysik Potsdam (AIP), Max-Planck-Institut f\"ur Astronomie (MPIA Heidelberg), Max-Planck-Institut f\"ur Astrophysik (MPA Garching), Max-Planck-Institut f\"ur Extraterrestrische Physik (MPE), National Astronomical Observatories of China, New Mexico State University, New York University, University of Notre Dame, Observat\'ario Nacional / MCTI, The Ohio State University, Pennsylvania State University, Shanghai Astronomical Observatory, United Kingdom Participation Group, Universidad Nacional Aut\'onoma de M\'exico, University of Arizona, University of Colorado Boulder, University of Oxford, University of Portsmouth, University of Utah, University of Virginia, University of Washington, University of Wisconsin, Vanderbilt University, and Yale University.

%%%%%%%%%%%%%%%%%%%%%%%%%%%%%%%%%%%%%%%%%%%%%%%%%%



\bibliographystyle{mnras}
\bibliography{QSO_prox} % if your bibtex file is called example.bib

%%%%%%%%%%%%%%%%%%%%%%%%%%%%%%%%%%%%%%%%%%%%%%%%%%

%%%%%%%%%%%%%%%%% APPENDICES %%%%%%%%%%%%%%%%%%%%%

\appendix

\section{Additional Composite Spectra of the the Quasar Proximity Effect on Metals}\label{apx:Stacks}
\begin{figure}
  \centering
  \includegraphics[width=.9\linewidth]{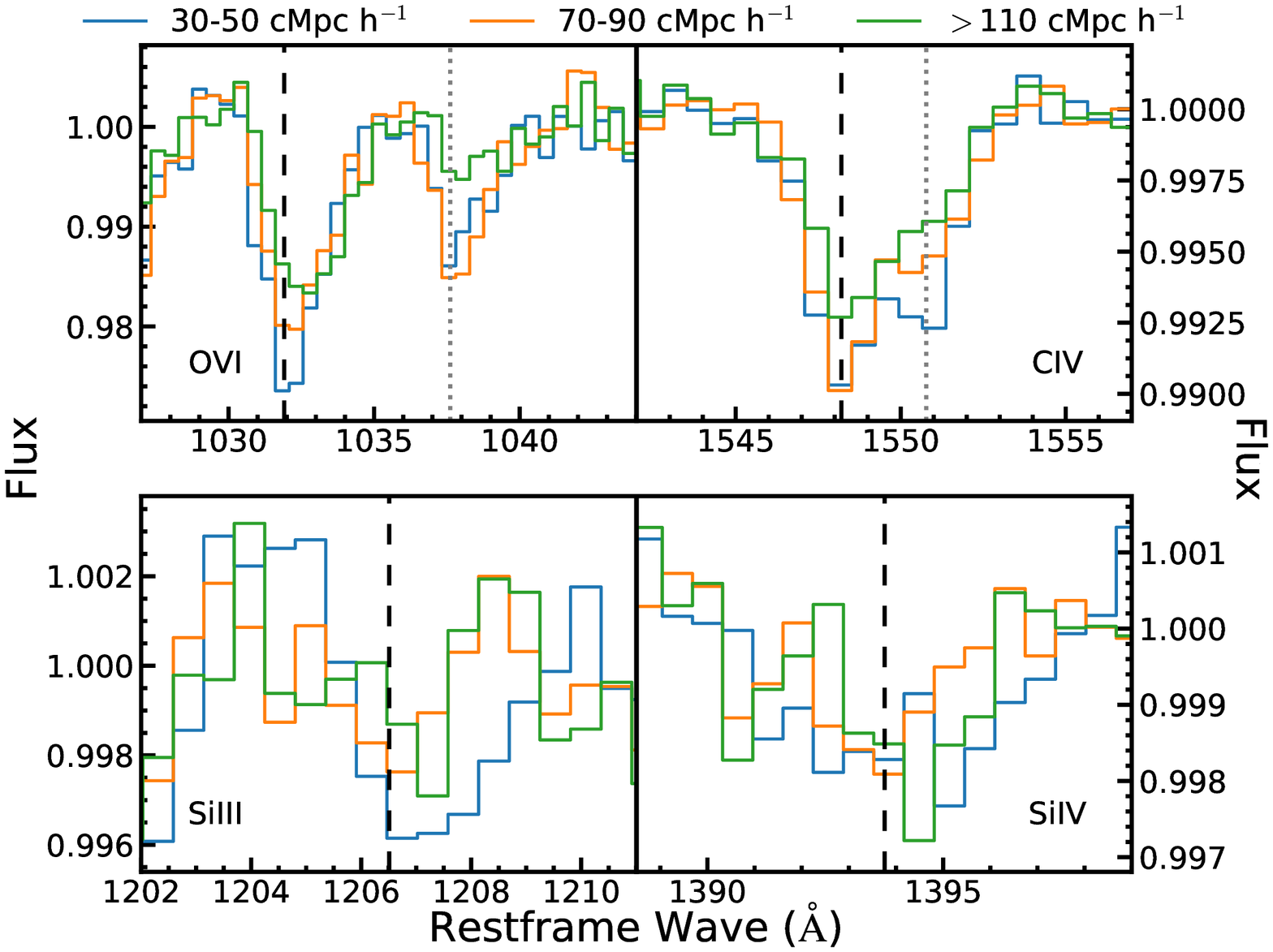}
  \caption{
  Additional (to \autoref{fig:IGM_stacks}) median composite spectrum of IGM ($2.4 <z<3.1$) absorption metal line for quasar proximity. The dashed black line is the wavelength of the metal line analyzed, and the gray dotted line (where applicable) is the location of the weaker line of the doublet.}
  \label{fig:IGM_stacks_add}
\end{figure}

\begin{figure}
  \centering
  \includegraphics[width=.85\linewidth]{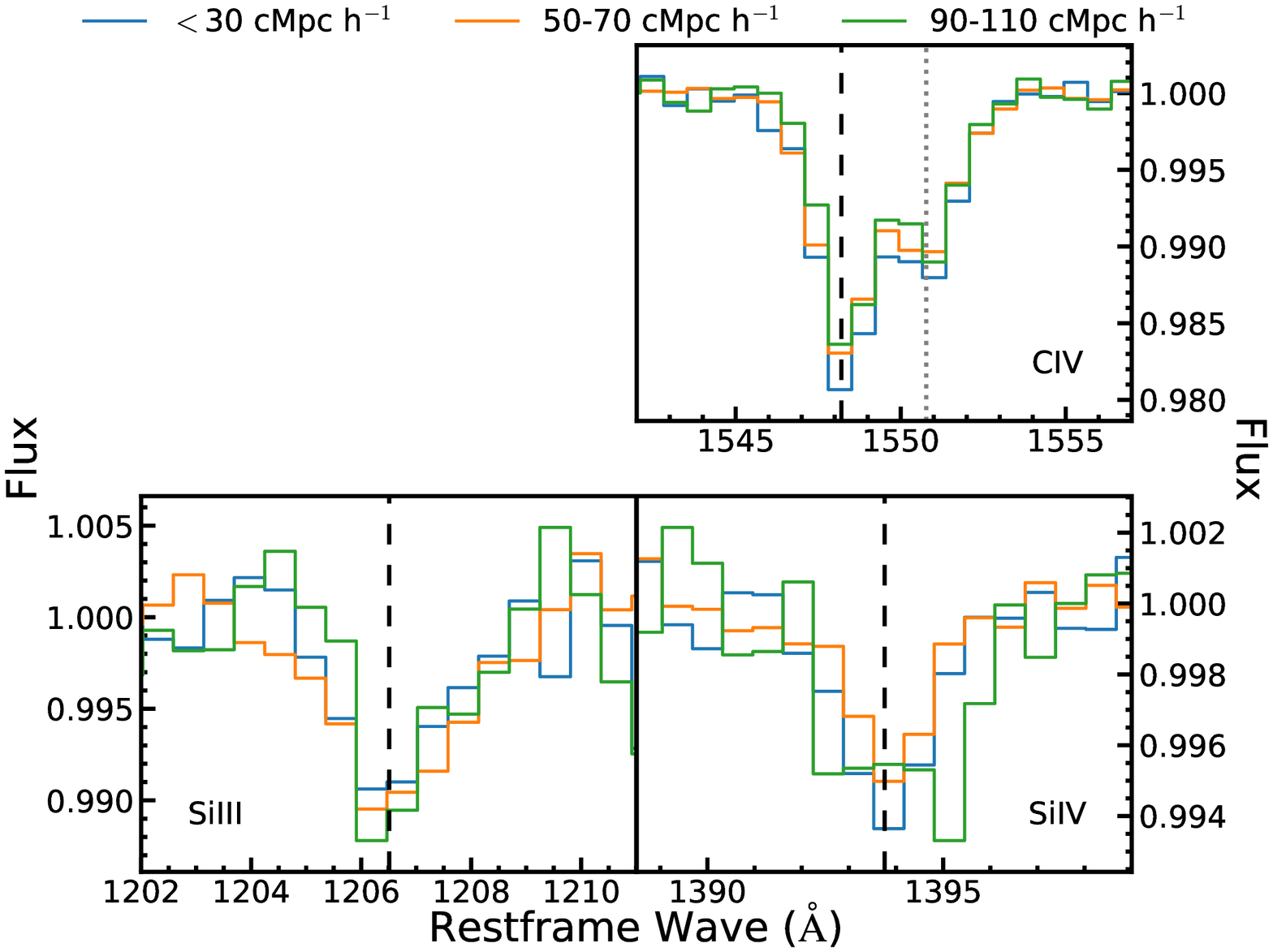}
   \includegraphics[width=.85\linewidth]{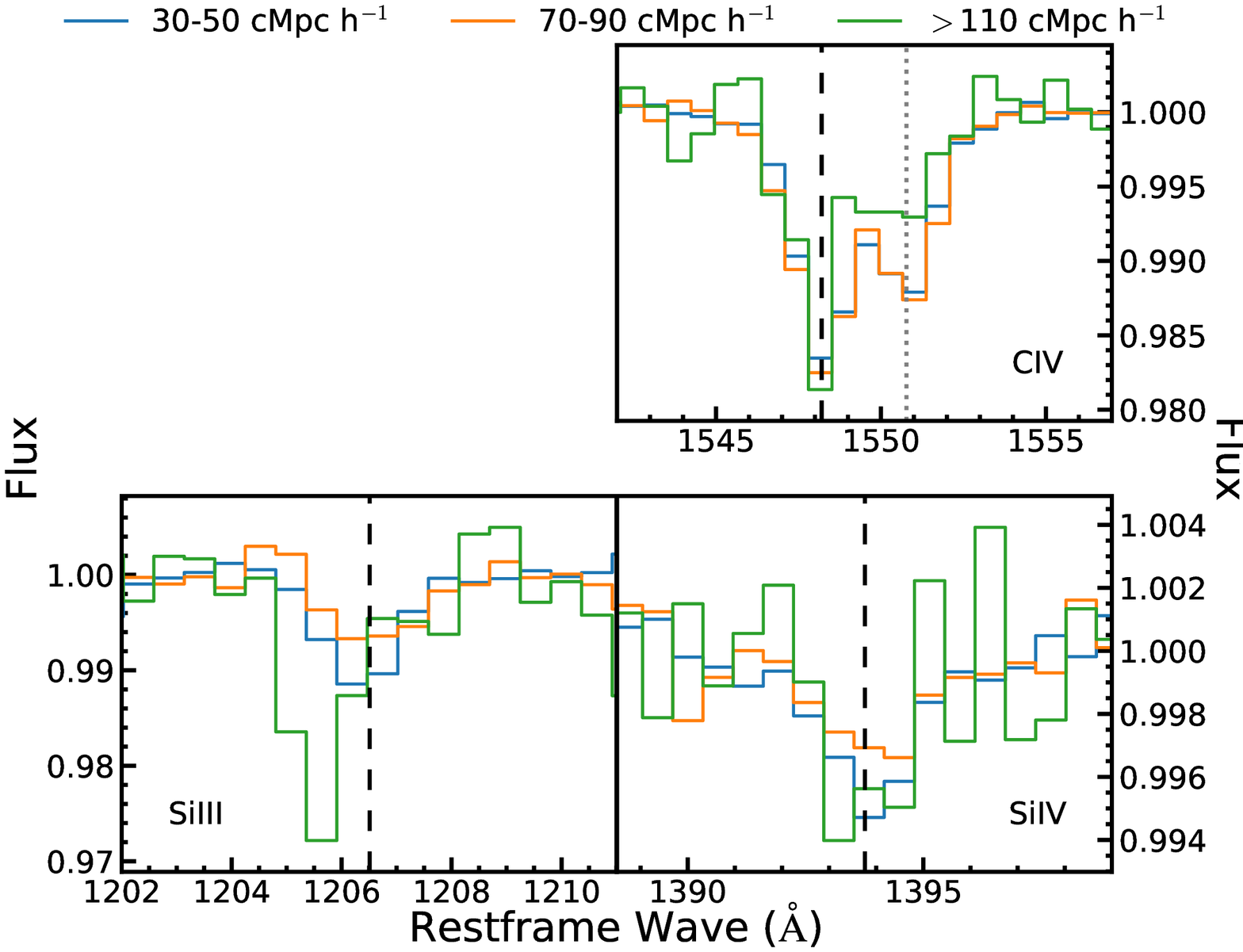}
  \caption{Same as Figures \ref{fig:IGM_stacks} and \ref{fig:IGM_stacks_add}, but for low-$z$ ($2.05<z<2.4$) IGM absorption composite spectra of various quasar proximity.}
  \label{fig:IGM_lz_stack}
\end{figure}

\begin{figure}
  \centering
  \includegraphics[width=.85\linewidth]{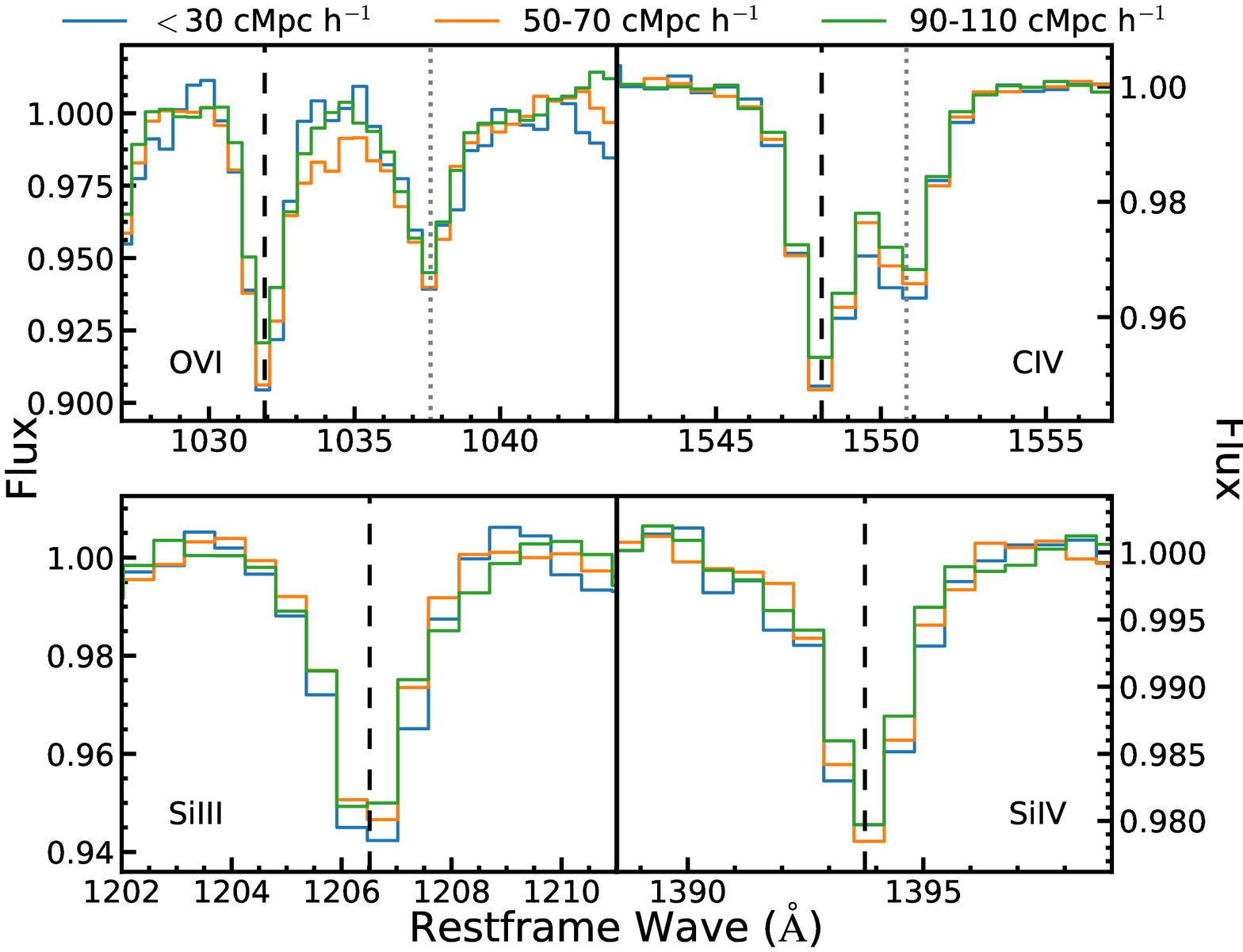}
  \includegraphics[width=.85\linewidth]{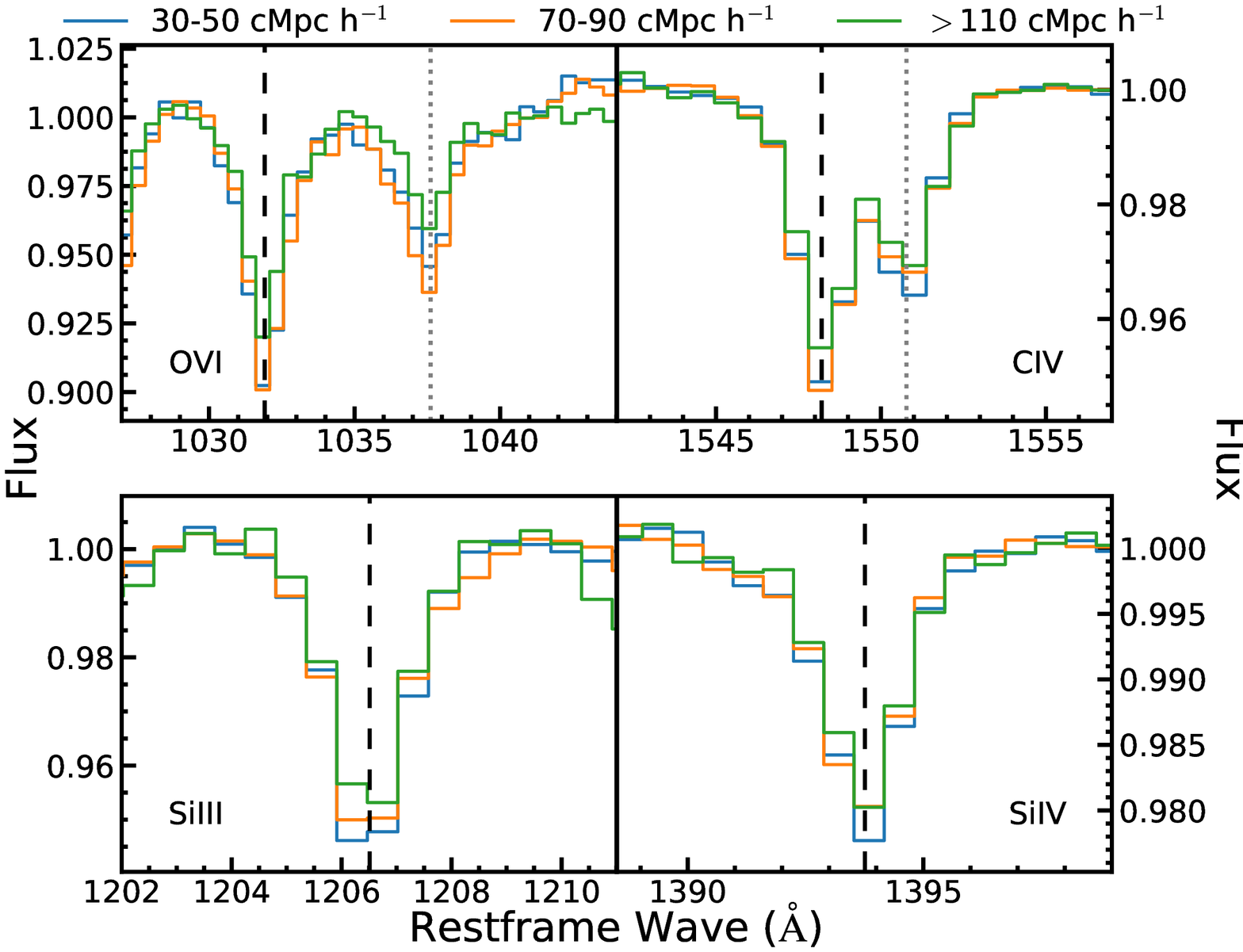}
  \caption{Same as Figures \ref{fig:IGM_stacks} and \ref{fig:IGM_stacks_add}, but for Mixed CGM ($2.4 <z<3.1$) absorption composite spectra of various quasar proximity.}
  \label{fig:CGM_stack}
\end{figure}

\begin{figure}
  \centering
  \includegraphics[width=.85\linewidth]{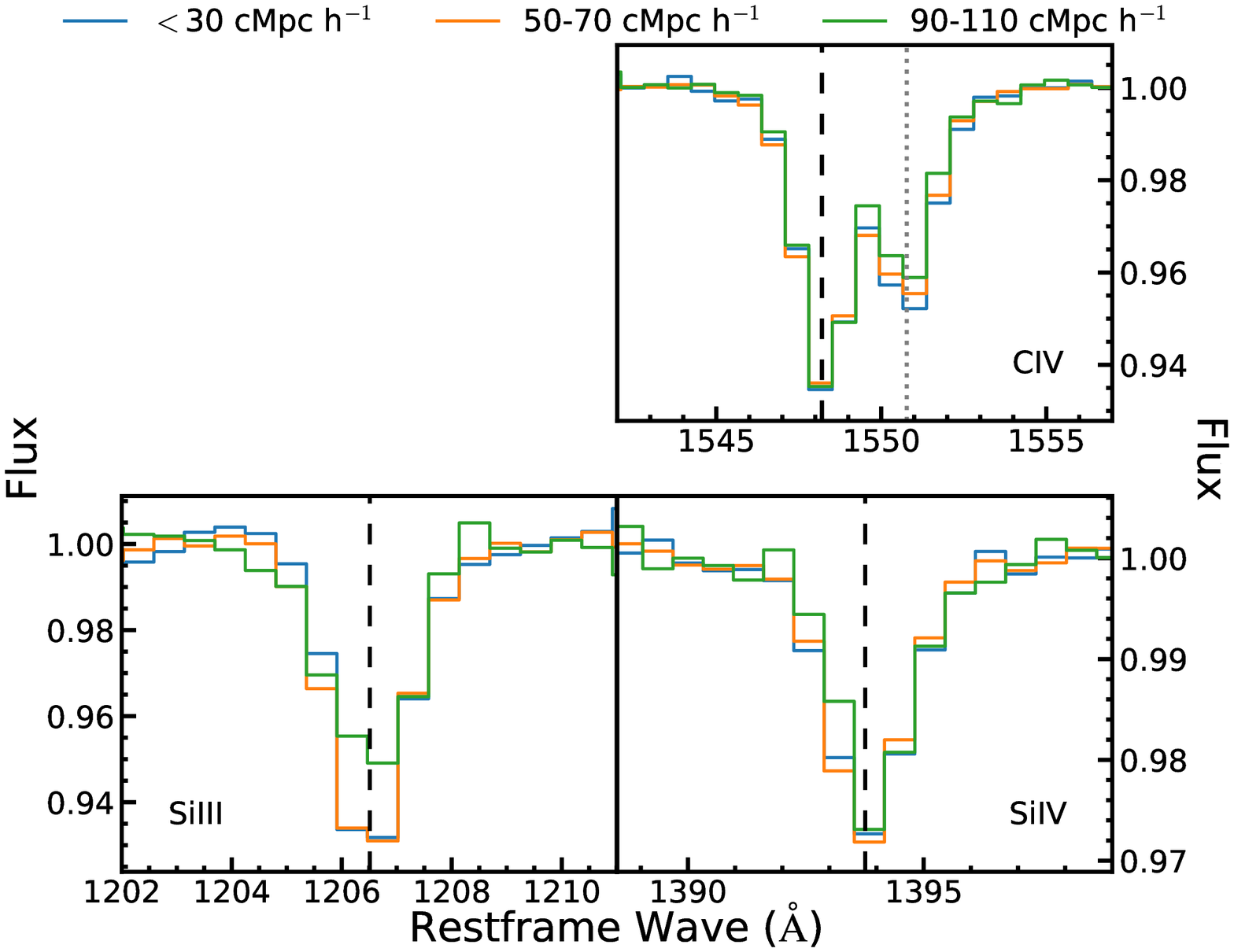}
  \includegraphics[width=.85\linewidth]{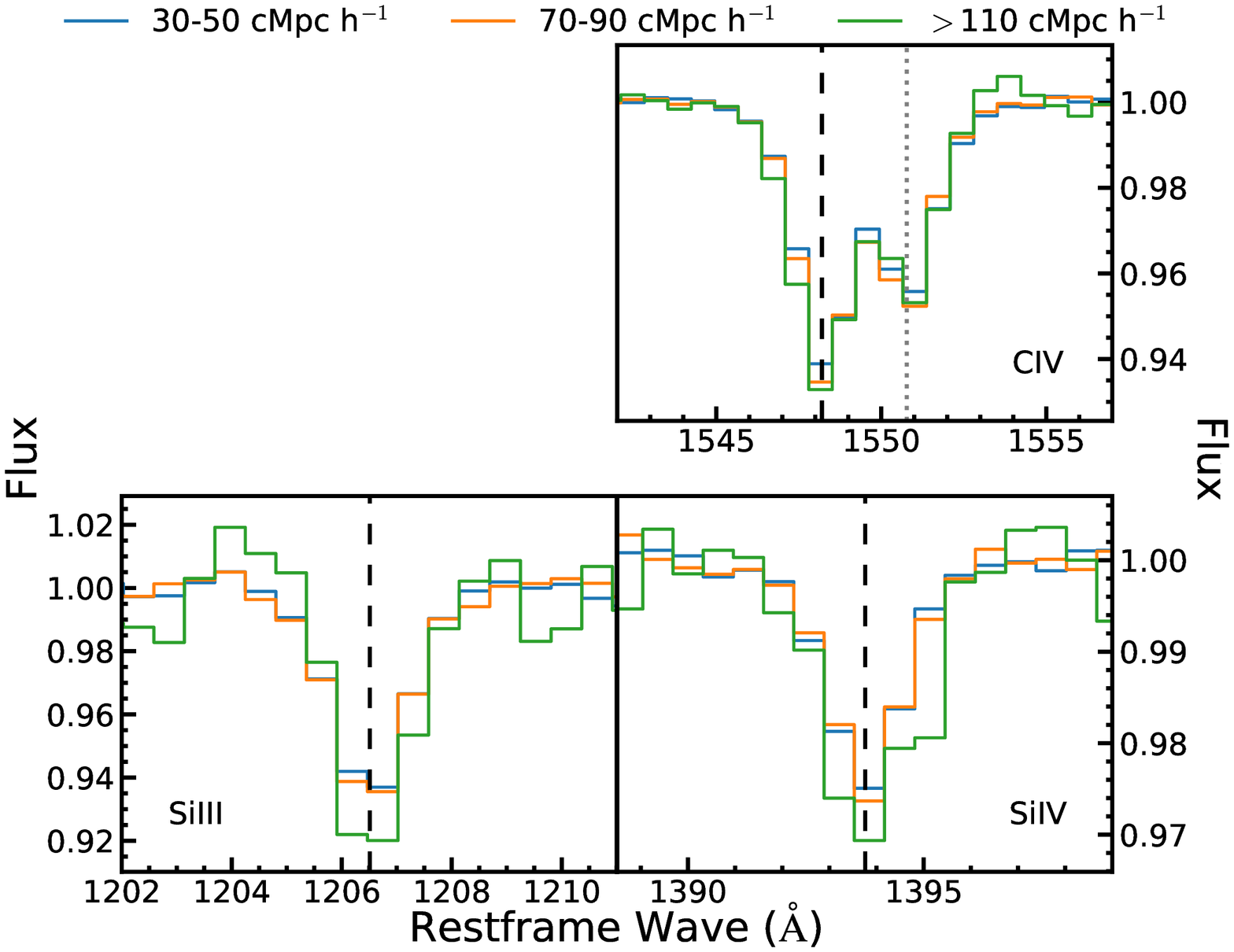}
  \caption{Same as Figures \ref{fig:IGM_stacks} and \ref{fig:IGM_stacks_add}, but for low-$z$ ($2.05<z<2.4$) Mixed CGM absorption composite spectra of various quasar proximity.}
  \label{fig:CGM_lz_stack}
\end{figure}

\begin{figure}
  \centering
  \includegraphics[width=.85\linewidth]{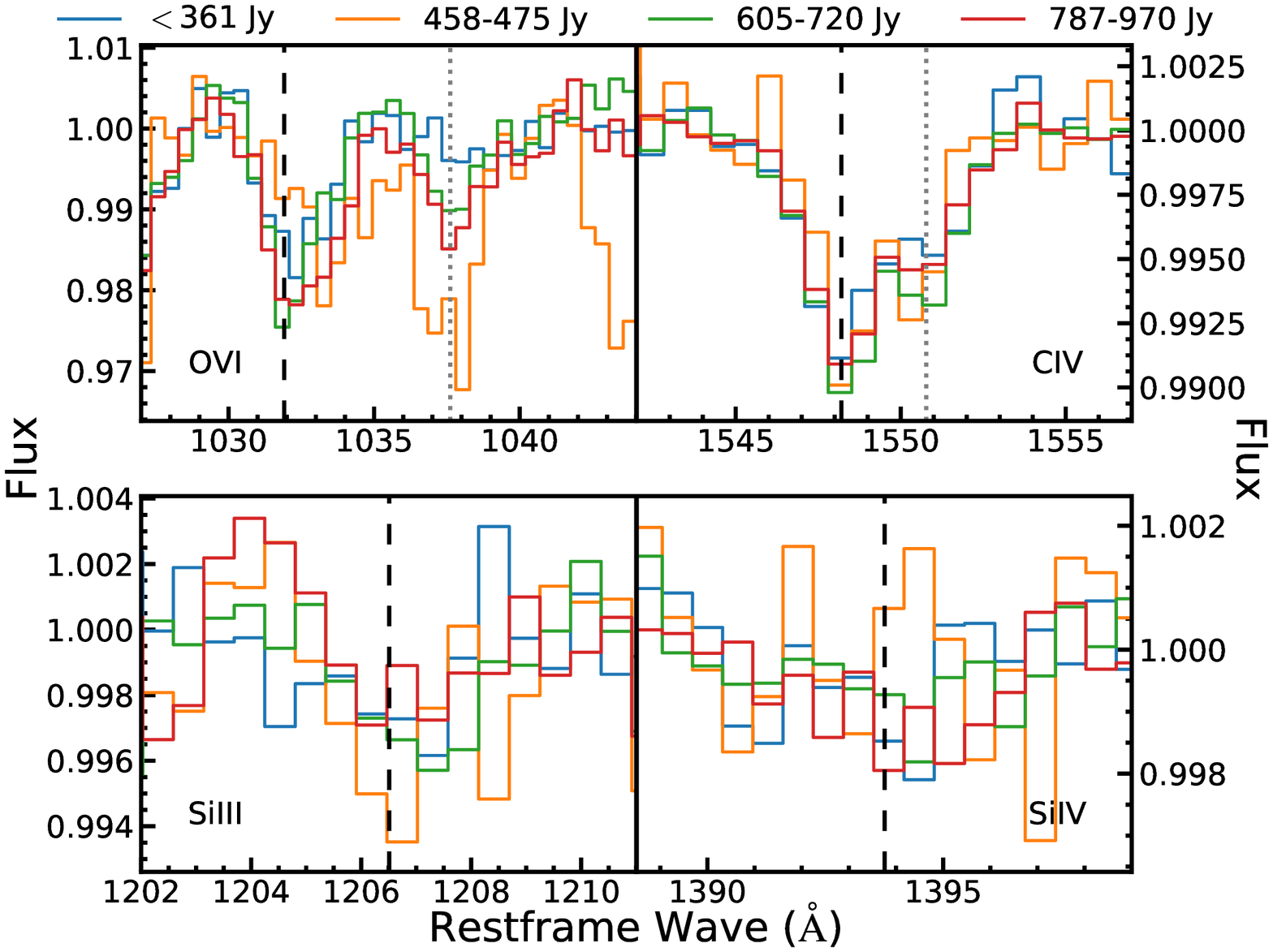}
  \includegraphics[width=.85\linewidth]{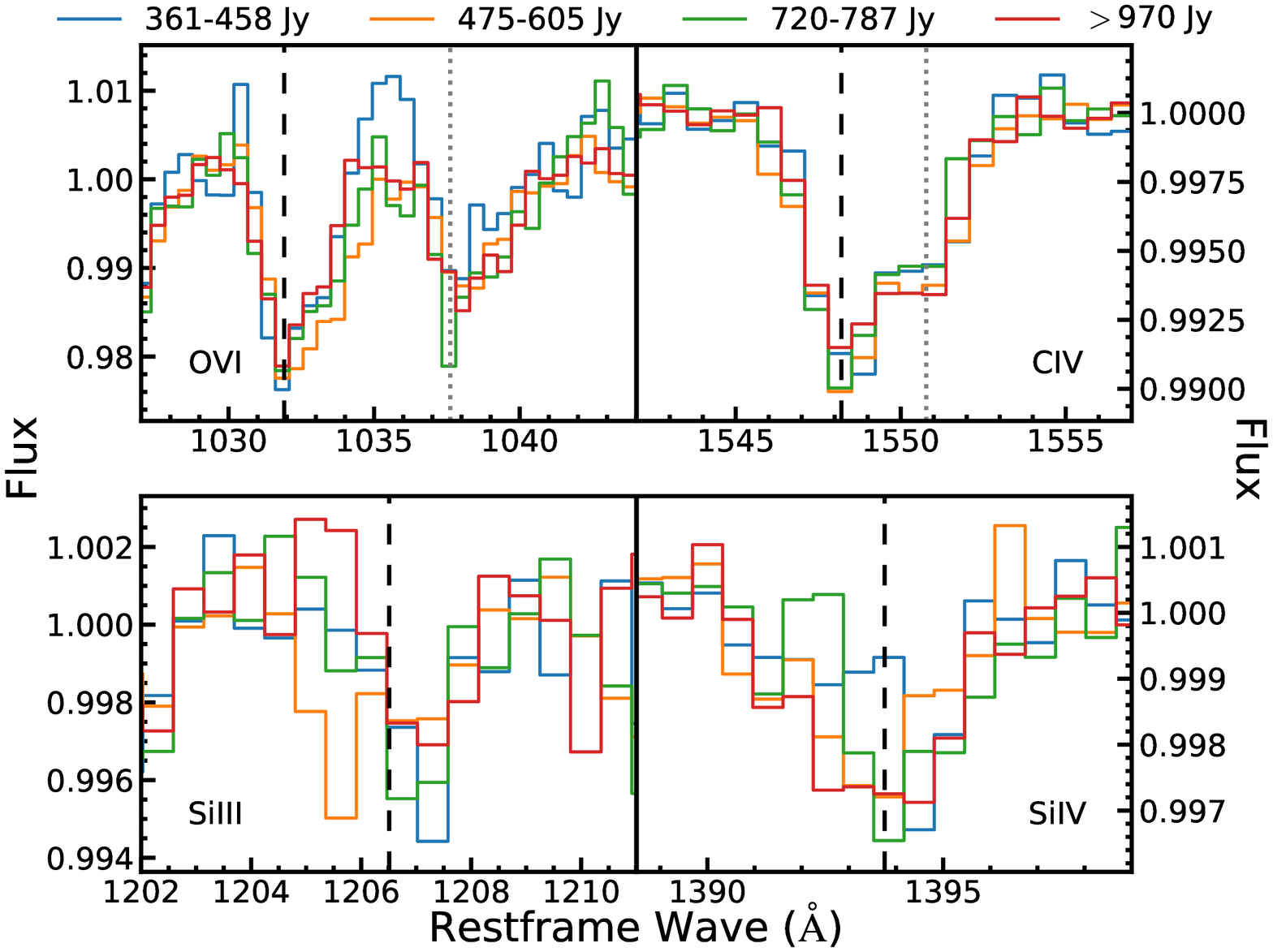}
  \caption{Same as Figures \ref{fig:IGM_stacks} and \ref{fig:IGM_stacks_add}, but for IGM absorption composite spectra of integrated flux in SDSS-r.}
  \label{fig:IGM_Int_stack}
\end{figure}

Figures \ref{fig:IGM_stacks_add} - \ref{fig:IGM_Int_stack} show the composite spectra for all lines (excluding those shown in \autoref{fig:IGM_stacks}) analyzed in this work. 

%%%%%%%%%%%%%%%%%%%%%%%%%%%%%%%%%%%%%%%%%%%%%%%%%%

% Don't change these lines
\bsp	% typesetting comment
\label{lastpage}
\end{document}